\documentclass[aps,prd,reprint,twocolumn,nofootinbib,
superscriptaddress]{revtex4-2}

\usepackage{graphicx}
\usepackage{amsmath}
\usepackage{amssymb}
\usepackage{bm}
\usepackage{placeins}
\usepackage{titlesec}
\usepackage[caption=false]{subfig}

\makeatletter
\titleclass{\subsubsubsection}{straight}[
  \subsubsection]
\newcounter{subsubsubsection}[subsubsection]
\renewcommand\thesubsubsubsection{\thesubsubsection .\arabic{subsubsubsection}}
\titleformat{\subsubsubsection}[runin]
  {\normalfont\normalsize\bfseries}{\thesubsubsubsection}{1em}{}
\titlespacing*{\subsubsubsection}{0pt}{3.25ex plus 1ex minus .2ex}{1em}

\usepackage{capt-of}
\usepackage{silence}
\usepackage{hyperref}

\begin{document}

\title{Primordial Tensor Signatures and Gravitational Wave Constraints in Lorentz Violating Inflation with Non-Canonical Kinetics}

\author{Dibash Plaban Kalita}
\email{24dr0070@iitism.ac.in}

\author{Shreya Banerjee}
\email{shreya@iitism.ac.in}

\affiliation{Department of Physics, Indian Institute of Technology (Indian School of Mines), Dhanbad, India}

\begin{abstract}
    We investigate primordial tensor perturbations in a Lorentz violating inflationary framework with non-canonical scalar field dynamics. Using an Einstein-\AE{}ther type with a higher order kinetic term $K_3X^2$, we derive the tensor propagation speed and its dependence on an effective Lorentz violating coupling. We consider an analytically tractable potential free background and study two representative couplings, $q_k=\lambda_T\phi_k^2$ and $q_k=\lambda_T\phi_k^4$. The resulting tensor spectral index $n_T$ and its running $\alpha_T$ satisfy distinct consistency relations, providing potential signatures of Lorentz violation in the primordial gravitational wave spectrum. We derive corresponding constraints from the tensor tilt and running and compare them with the stringent late-time bound on the gravitational wave speed from GW170817/GRB170817A. We emphasize that translating this bound into primordial constraint requires an assumption about the evolution of tensor sector coupling between inflation and the late Universe. Our results demonstrate that primordial tensor observables can provide a complementary probe of Lorentz violation and its interplay with non-canonical inflationary dynamics.
\end{abstract}

\maketitle
\section{INTRODUCTION}
Inflation provides a compelling framework for describing the dynamics of the early universe and offers a natural mechanism for generating the primordial perturbations that seed the observed large scale structure \cite{Guth1981,Linde1982,AlbrechtSteinhardt1982,MukhanovChibisov1981}. In addition to scalar fluctuations, inflation generically produces a background of primordial gravitational waves. While the scalar sector has been extensively constrained by observations of cosmic microwave background (CMB), the tensor sector provides an independent window into the physics of the early universe. The amplitude and spectral properties of primordial tensor perturbations can probe the energy scale of inflation and provide information about the gravitational dynamics operating in the early universe \cite{Baumann2009, Planck2020}. In particular, the tensor spectral index, its running, and the propagating properties of gravitational waves can provide useful discriminants among different inflationary and gravitational scenarios. The continuing improvement in CMB polarisation measurements and the development of future gravitational-wave observatories therefore motivate a systematic investigation of primordial tensor modes beyond the standard inflationary framework \cite{Baumann2009,Planck2020}.

Although General Relativity (GR) has been remarkably successful in describing gravitational phenomena over a wide range of scales, its validity at the highest energy scales remains an open theoretical question. Local Lorentz invariance, one of the fundamental symmetries underlying GR, may not necessarily remain exact in such a regime, motivating Lorentz-violating extensions of gravity \cite{Mattingly2005, Kostelecky2004}. Lorentz violation has been investigated in several approaches to quantum gravity and effective field theory, including the standard-model extension and theories containing a preferred timelike direction \cite{Mattingly2005,Kostelecky2004,Jacobson2007,KosteleckySamuel1989,CarrollLim2004,JacobsonMattingly2001}. Cosmology provides a particularly natural setting for such theories because the homogeneous Universe already selects a preferred temporal direction. During inflation, Lorentz-violating effects can modify both the background dynamics and the propagation of cosmological perturbations, including gravitational waves \cite{JacobsonMattingly2004,KannoSoda2006}.

The observational relevance of gravitational wave propagation was highlighted by the multimessenger observation of GW170817 and GRB170817A, which placed extremely stringent constraints on the deviation of gravitational wave speed from the speed of light at late times \cite{Abbott2017GW170817,Abbott2017GRB170817A}. These constraints strongly restrict broad classes of modified gravity theories, while still leaving open the possibility that Lorentz-violating effects were relevant during the much earlier inflationary epoch \cite{KannoSoda2006,Oost2018,Baker2017,SaksteinJain2017}. This motivates models in which Lorentz violation is dynamically important at high energies but becomes suppressed as the Universe evolves towards the low-energy regime.

Einstein-\AE{}ther gravity provides a well-defined generally covariant realization of Lorentz-violating gravity through a dynamical unit time-like vector field, \AE{}ther, which selects a preferred local frame \cite{Jacobson2007,JacobsonMattingly2001,JacobsonMattingly2004}. Its phenomenology has been extensively studied in cosmology, post-Newtonian gravity, stability analyses, and gravitational wave propagation \cite{JacobsonMattingly2001,JacobsonMattingly2004,FosterJacobson2006}. Importantly, the tensor, vector and scalar modes depend on different combinations of the \AE{}ther couplings, making the tensor sector complementary to the background and other observational constraints \cite{JacobsonMattingly2001,JacobsonMattingly2004,FosterJacobson2006}. Inflationary applications further show that Lorentz violation can modify the dynamics of chaotic inflation and potentially evolve toward the conventional slow-roll regime \cite{KannoSoda2006}, while perturbative analyses indicate that scalar and vector stability can impose additional restrictions on coupling parameters \cite{SolomonBarrow2014}.

A further extension is obtained by allowing the scalar sector to possess non-canonical kinetic dynamics. The canonical form $P(X,\phi)=X-V(\phi)$ is not the only possible description of inflation, and higher-order operators in the kinetic invariant $X$ can arise naturally in effective descriptions of high energy physics. Non-canonical inflation and k-inflation provide well-established frameworks for investigating such effects \cite{ArendarmizPicon1999,GarrigaMukhanov1999,ArendarmizPicon2001}. In general $P(X,\phi)$ theories, the kinetic structure modifies the background evolution, scalar sound speed, and primordial observables, including the scalar spectrum and non-Gaussianity \cite{GarrigaMukhanov1999,ArendarmizPicon2001,Chen2010}. Such corrections may therefore become particularly relevant during inflation. 

In this work, we consider the simplest non trivial non canonical extension, $P(X,\phi)=X+K_3 X^2-V(\phi)$, with $X=-1/2\nabla_{\mu}\phi\nabla^{\mu}\phi$. The $K_3 X^2$ term represents the leading higher-order kinetic correction, with the canonical limit recovered for $K_3=0$.Its presence modifies the scalar kinetic response through $P_X=1+2K_3X$ and $P_{XX}=2K_3$, and consequently affects the inflationary background and scalar perturbations \cite{ArendarmizPicon1999,GarrigaMukhanov1999,ArendarmizPicon2001}.

The combination of non-canonical scalar dynamics with Lorentz-violating gravity is particularly interesting because the two sectors affect different parts of the gravitational system. The $K_3 X^2$ interaction modifies the inflaton dynamics and hence the evolution of $H$ and $\phi$, while the \AE{}ther sector modifies the tensor kinetic structure and gravitational wave propagation. When the effective \AE{}ther coupling depends on the inflaton, these effects become dynamically connected: changes in the scalar evolution can induce changes in the \AE{}ther coupling and consequently in the tensor kinetic coefficient and propagation speed. The primordial tensor spectrum can therefore carry signatures of both the non-canonical scalar sector and Lorentz-violating gravity.

An important feature of the Einstein-\AE{}ther gravity is that the coupling combination governing the background, $\beta_1+2\beta_2+\beta_3$, differs from that entering the tensor sector, $\beta_T=\beta_1+\beta_3$ \cite{JacobsonMattingly2001,JacobsonMattingly2004,FosterJacobson2006}. Thus, background observations do not uniquely determine the tensor propagation properties, making primordial gravitational waves a complementary probe of Lorentz violating couplings \cite{JacobsonMattingly2004,Oost2018,FosterJacobson2006}.

To investigate the phenomenological consequences of a field dependent Lorentz-violating coupling while retaining analytical control, we consider the representative class $\beta(\phi)=\xi \phi^p$ and focus on two lowest non-trivial even powers, $p=2,\quad 4$. The quadratic coupling, $\beta(\phi)=\xi \phi^2$, represents the minimal analytic extension of a constant Einstein-\AE{}ther coupling and preserves the $\phi\rightarrow -\phi$ symmetry. The quartic coupling, $\beta(\phi)=\xi \phi^4$, provides the next non trivial even power realization and allow us to examine how the tensor sector changes when the field dependence of the Lorentz violating coupling becomes stronger. Importantly, these two choices lead to qualitatively distinct primordial tensor spectra and consistency relations, including, different predictions for the tensor tilt and its running. They therefore provide a simple but non trivial framework for assessing whether primordial gravitational wave observables can discriminate between different functional forms of Lorentz violation.  

In particular, we derive the tensor quadratic action, from which the effective tensor kinetic coefficient and propagation speed are obtained as $\gamma=1-16\pi G(\beta_1+\beta_3)$ and $c_T^2=1/\gamma$. The field dependence of the \AE{}ther coupling consequently leads to a time-dependent tensor normalization and propagation speed. We then calculate the primordial tensor spectrum and investigate its spectral index, $n_T$ and $\alpha_T$ with particular emphasis on their dependence on $K_3, \xi$ and functional form of the coupling.

Our central objective is to determine whether the combined effects of non canonical scalar kinetics and dynamically evolving Lorentz violation can generate distinctive primordial gravitational wave signatures while maintaining a theoretically consistent tensor sector. In particular, we examine how the $K_3X^2$ correction influences the inflationary background and, through the field dependent \AE{}ther coupling, affects the tensor kinetic coefficient, propagation speed, tensor tilt and its running. The resulting chain $K_3X^2\rightarrow$modified scalar dynamics$\rightarrow H(t),\phi(t)\rightarrow \gamma(t),c_T\rightarrow P_T(k)$, provides a direct connection between high energy scalar dynamics and primordial gravitational wave observables.

The paper is organized as follows. Section II introduces the Einstein-\AE{}ther theory with the non canonical scalar sector and derives the background equations. Section III discusses the inflationary dynamics for the quadratic and quartic forms of the field dependent \AE{}ther coupling. Section IV derives the quadratic tensor action and the primordial tensor spectrum, including its spectral index and running. Section V discusses theoretical and observational constraints on the model parameters. Section VI summarizes the main results and their implications for primordial gravitational wave phenomenology.

\section{Lorentz violation with non-canonical kinetic term}

\subsection{Einstein \AE{}ther action with non-canonical scalar}

We consider an inflationary extension of Einstein \AE{}ther gravity in which the scalar field possesses a non canonical kinetic sector and couples to the \AE{}ther through field dependent interaction functions. This framework allows us to investigate the inter play between Lorentz violating gravitational dynamics and higher order scalar kinetic corrections during inflation.

We work within Einstein \AE{}ther theory where Lorentz symmetry is spontaneously broken by the introduction of a unit timelike vector field  $u^{\mu}$ having the expectation value
\begin{equation}
    <0|u_\mu u^\mu|0>=-1
\end{equation}
The mechanism giving rise to this expectation value is discussed in \cite{KosteleckySamuel1989}. The coupling of such a vector field with gravity during inflation has been investigated previously in \cite{KannoSoda2006}. It has been shown that such coupling can have a significant impact on the inflationary dynamics of the universe.

To distinguish the different physical contributions, we express the total action as 
\begin{equation}                  S=S_{\rm{EH}}+S_{\mathrm{\text{\AE}}}+S_{\mathrm{constraints}}+S_{\rm{\phi}}
\end{equation}
Here, the Einstein-Hilbert term $S_{\rm{EH}}$ governs the standard gravitational dynamics, while the \AE{}ther action  $S_{\rm{\text{\AE}}}$ introduces Lorentz-violating effects mediated by the field dependent couplings $\beta_i(\phi)$ \cite{KannoSoda2006,JacobsonMattingly2001,Avelino_2009}. The constraint action $S_{\rm{constraint}}$ incorporates the Lagrange multiplier $\lambda$ to ensure the unit timelike condition, and the scalar sector $S_\phi$ is governed by the generalised function $P(\phi, X)$, allowing non-canonical kinetic effects driving inflation \cite{1999PhLB..458..209A}. We extend this theoretical framework to examine observational signatures through tensor perturbations. The complete action takes the form

\begin{equation}\label{action}
\begin{aligned}
    S = \int d^{4} x \sqrt{-g} \bigg[ & \frac{R}{16 \pi G} - \beta_{1}(\phi) \nabla^{\mu} u^{\nu} \nabla_{\mu} u_{\nu} \\
    & - \beta_{2}(\phi) \nabla^{\mu} u^{\nu} \nabla_{\nu} u_{\mu} \\
    & - \beta_{3}(\phi) (\nabla_{\mu} u^{\mu})^{2} \\
    & - \beta_{4}(\phi) u^{\mu} u^{\nu} \nabla_{\mu} u^{\alpha} \nabla_{\nu} u_{\alpha} \\
    & + \lambda (u^{\mu} u_{\mu} + 1) - P(\phi,X) \bigg]
\end{aligned}
\end{equation}
Here $\beta_i(\phi)$ are the field-dependent coupling functions, and $\lambda$ is the Lagrange multiplier enforcing the unit norm constraint. As $u^{\mu}$ is dimensionless, the dimension of $\beta_i$ is mass squared, or $\sqrt{\beta_i}$ has the dimension of mass.
$P(\phi, X)$ is the general Lagrangian density depending  on the scalar and its kinetic form
\begin{equation}
    X=\frac{1}{2\mathcal{N}^2}(\nabla\phi)^2
\end{equation}
 To capture the leading departure from canonical scalar dynamics, we consider the simplest non-canonical kinetic function
 $P(\phi,X)$ given by \cite{Li_2012}
\begin{equation}\label{P form}
    P(\phi,X)=K_2X+K_3X^2-V(\phi)
\end{equation}
The scalar field $\phi$ can be rescaled to set $K_2=1$, eliminating a redundant constant while preserving the physics \cite{Li_2012}. The parameter $K_3$ controls the relative importance of the higher-order kinetic contribution and reduces to a canonical form in the limit $X\to0$, while finite $K_3$ allows us to investigate the impact of kinetic corrections on the inflationary background and primordial perturbations\cite{Li_2012,GarrigaMukhanov1999}. Current observational evidence does not constrain $K_3$ \cite{Li_2012}, so based on MCMC sampling, we adopt $K_3<10^{20}$ \cite{Li_2012}. With this non-canonical kinetic term in the Lagrangian, we explore how Lorentz violation and non-canonical scalar dynamics affect the background and tensor perturbations.

Now, from Eq.\eqref{P form}, we have
\begin{equation}
\begin{aligned}
    &P,_X=1+2K_3X\\
    &P,_{XX}=2K_3
\end{aligned}
\end{equation}
Since $K_3>0$ and $X>0$ during inflation,
\begin{equation}
    1+2K_3X>0
\end{equation}
which ensures that
\begin{equation}
    P,_X>0
\end{equation}

The scalar sound speed is given by
\begin{equation}
    c_s^2= \frac{P_X}{P_X+2XP_{XX}} = \frac{1+2K_3X}{1+6K_3X}.
\end{equation}

As $K_3>0$ and $X>0$, both the numerator and denominator remain positive, yielding
\begin{equation}
    0<c_s^2\leq 1
\end{equation}
Therefore, the scalar perturbation remains stable throughout the inflationary evolution.

Now, we consider the chaotic inflationary scenario \cite{LINDE1983177} and determine how strongly we can observe the effects of our new model. The preferred frame that the vector $u^\mu$ determines may be different from the CMB frame. However, these frames align during expansion. This is shown in the reference \cite{KannoSoda2006}. Therefore, we consider homogeneous and isotropic spacetime,
\begin{equation}
    ds^2=-\mathcal{N}(t)^2dt^2+e^{2\alpha(t)}\delta_{ij}dx^idx^j
\end{equation}
where $\mathcal{N}$ is the lapse function and $\alpha=ln a(t)$ determines the scale factor of the universe. Since the vector field $u_\mu$ is constrained to have a timelike direction, we therefore have to take \cite{KannoSoda2006}
\begin{equation}\label{unit norm form}
    u^\mu=\left(\frac{1}{\mathcal{N}},0,0,0\right)
\end{equation}
The cosmological implications of preferred timelike directions have been discussed in \cite{CarrollLim2004}. By calculating the action from the constraint of the timelike vector field, the Lagrangian for the Lorentz-violating factor with the non-canonical kinetic term is
\begin{equation}\label{eq1}
    L=\frac{e^{3\alpha}}{\mathcal{N}}\left[-\frac{3}{8\pi G}(1+8\pi G\beta)\dot\alpha^2+\mathcal{N}^2P(\phi,\nabla\phi)\right]
\end{equation}
where $\beta(\phi)=\beta_1+2\beta_2+\beta_3$.

Only this particular combination of $\beta_1,\beta_2$ and $\beta_3$ contributes to the dynamics. The parameter $\beta_4$ doesn't make any contribution as it vanishes under the ansatz (\ref{unit norm form}). Thus, $\beta(\phi)$ determines the effective coupling that determines the strength of Lorentz-violating effects. For convenience, we define a dimensionless derivative $R'$ 
\begin{equation}
    \dot{R}=\frac{dR}{d\alpha}\frac{d\alpha}{dt}\equiv R'\frac{d\alpha}{dt}
\end{equation}
where we use throughout in our work.

\subsection{Homogeneous background equations}
Varying the reduced action \eqref{eq1} with respect to the lapse function $\mathcal{N}$, the scale factor $\alpha$, and the scalar field $\phi$ respectively, yields the homogeneous background equations of motion.
\begin{equation}\label{eq2}
    3\left(1+\frac{1}{8\pi G\beta}\right)H^2=\frac{2XP,_X-P}{\beta}
\end{equation}
\begin{equation}\label{eq3}
    \frac{\beta'}{\beta}+\left(1+\frac{1}{8\pi G\beta}\right)\frac{H'}{H}+\frac{XP,_X}{\beta H^2}=0
\end{equation}
\begin{equation}\label{eq4}
    \frac{d}{d\alpha}(P,_{\phi'})+\left(3-\frac{H'}{H}\right)P,_{\phi'}+3H^2\beta,_{\phi}-P,_{\phi}=0
\end{equation}
where $\beta,_\phi$, $P,_\phi$ represent the derivative with respect to $\phi$ and $P,_X$ represents the derivative with respect to X. $H=\dot{\alpha}$ is taken as an independent variable. $\mathcal{N}=1$ is set after taking variations. Eq. \eqref{eq2} is the modified Friedmann equation with Lorentz-violating effects, Eq. (\ref{eq3}) governs the evolution of the effective coupling with the expansion of the universe, and Eq. (\ref{eq4}) is the modified Klein-Gordon equation for the inflaton field. These equations show that the \AE{}ther sector modifies the effective gravitational response through the combination $\beta=\beta_1+2\beta_2+\beta_3$. The scalar dynamics are simultaneously modified by the non-canonical kinetic contribution through $P,_X$. Thus, the background evolution is governed by the coupled evolution of $H,\phi$ and the effective Lorentz-violating coupling. Substituting $P(\phi,X)$ from Eq. \eqref{P form}, the equations of motion become
\begin{equation}\label{eq5}
    3\left(1+\frac{1}{8\pi G\beta}\right)H^2=\frac{1}{\beta} \left(\frac{1}{2} H^2 \phi '^2 +\frac{3K_3}{4} H^4 \phi '^4 +V(\phi)\right)
\end{equation}

\begin{equation}\label{eq6}
    \frac{\beta '}{\beta}+\left(1+\frac{1}{8 \pi G \beta}\right)\frac{H'}{H}+\frac1\beta\left(\frac12\phi'^2+K_3H^2\phi'^4\right) =0
\end{equation}

\begin{equation}\label{eq7}
\begin{aligned}
    \left(1+3 K_3 H^2 \phi '^2\right)\phi ''  +\left(3+\frac{H'}{H}\right)\phi '+ \\ 
     3\left(\frac{H'}{H}+1\right) K_3 H^2 \phi'^3 
      +3\beta_{,\phi} + \frac{V_{,\phi}}{H^2}=0
\end{aligned}
\end{equation}

Eqs.\eqref{eq5}-\eqref{eq7} show explicitly how the non-canonical kinetics and the \AE{}ther sector modify the background dynamics through $K_3$ and the effective coupling $\beta(\phi)$, respectively. The coefficient $K_3$ introduces the higher-order kinetic term $(\phi'^4)$ which modifies the scalar field dynamics and its energy density. Furthermore, the field-dependent scalar-\AE{}ther coupling $\beta(\phi)$ also acts as the driving force during inflation along with the scalar potential $V(\phi)$. Thus, both $K_3$ and $\beta(\phi)$ govern the overall background dynamics. These reduce to the standard chaotic inflation in the limit $\beta\to0$ and $K_3\to 0$.

The factor $\left(1+\frac{1}{8\pi G\beta}\right)$ plays an important role in determining the dynamics of the inflationary scenario. When $8\pi G \beta \gg 1$, Lorentz-violating effects are dominant, whereas for $8\pi G\beta \ll1$, the equations reduce to the standard general relativistic case \cite{KannoSoda2006}. This change of dynamics is defined by the critical value $\phi_c$
\begin{equation}\label{critical field value}
    8\pi G\beta(\phi_c)=1
\end{equation}
The universe needs an accelerated expansion of about $N=60$ e-foldings to explain inflation \cite{Liddle:2000cg}. Let $\phi_i$ be the initial value of the scalar field. When $\phi_c>\phi_i$, then the Lorentz-violation effects can be neglected throughout inflation. On the other hand, if $\phi_c<\phi_i$, the Lorentz-violating term becomes important and requires model-dependent modifications.

\subsection{Our models}
To illustrate the effect of Lorentz violation on inflationary dynamics, we investigate explicit models for the coupling function $\beta(\phi)$. So, to obtain explicit solutions and investigate phenomenological consequences of Lorentz violation during inflation, we consider two representative forms of the coupling function $\beta(\phi)$ such that it provides a convenient framework for exploring how Lorentz violation with non-canonical kinetic influence on inflationary dynamics \cite{KannoSoda2006}:

\begin{equation}\label{model1}
    \beta=\xi\phi^2 \qquad, \qquad V(\phi)=\frac12m^2\phi^2
\end{equation}
and
\begin{equation}\label{model2}
    \beta=\xi\phi^4 \qquad, \qquad V(\phi)=\frac12m^2\phi^2
\end{equation}
where $\xi$ and $m$ are constant model parameters. The purpose is not merely to consider two functional forms, but to determine how the field dependence of the Lorentz-violating interaction changes the inflationary background and, subsequently, the primordial tensor spectrum. 

We consider the same quadratic potential for both the coupling models so that any differences in the resulting inflationary dynamics can be attributed directly to the Lorentz-violating coupling function $\beta(\phi)$. For both models \eqref{model1} and \eqref{model2}, the critical field value defined in Eq. \eqref{critical field value}, is given, respectively, by
\begin{equation}
    \phi_c=\frac{M_{pl}}{\sqrt{8\pi \xi}}
\end{equation}
\begin{equation}
    \phi_c=\frac{M_{pl}^{\frac12}}{(8\pi\xi)^{\frac14}}
\end{equation}

As we know, in the standard case $\phi_i\sim 3M_{pl}$ approximately, so by using the condition $\phi_i>\phi_c$, we get $1/72\pi<\xi<1/16$ and $1/648\pi<\xi<1/324$ respectively for the models (\ref{model1}) and (\ref{model2}) to have the effects of Lorentz violation during inflation. Beyond these ranges, when $\phi_c>\phi_i$, Lorentz-violating effects are not significant.

\subsection{Lorentz-violating inflationary regime} \label{LV violating subsection}
To investigate inflationary dynamics where Lorentz-violating effects dominate ($8\pi G\beta(\phi)\gg1$), we consider the initial condition $\phi_i>\phi_c$. In this regime, both the coupling function $\beta(\phi)$ and the scalar potential $V(\phi)$ become significant to the evolution of the scalar field. In the limit $8\pi G\beta(\phi)\gg1$, Eqs. \eqref{eq5}-\eqref{eq7} reduce to
\begin{equation}\label{eq8}
     H^2 =\frac{1}{3 \beta}\left(\frac{1}{2}H^2\phi '^2 +\frac{3K_3}{4}H^4 \phi '^4 +V(\phi)\right)
\end{equation}
\begin{equation}\label{eq9}
    \frac{\beta '}{\beta}+\frac{H'}{H}+\frac1\beta\left(\frac12\phi'^2+K_3H^2\phi'^4\right)=0
\end{equation}
\begin{equation}\label{eq10}
\begin{aligned}
       \left(1+3 K_3 H^2 \phi '^2\right)\phi '' &+\left(3+\frac{H'}{H}\right)\phi' \\
       &+4\left(\frac{H'}{H}+1\right) K_3 H^2 \phi'^3\\
       &+3\beta,_\phi +\frac{V,_\phi}{H^2}=0 
\end{aligned}
\end{equation}

Eqs. \eqref{eq8}-\eqref{eq10} show that, in the regime where Lorentz-violating effects are dominant, the background evolution is primarily governed by the \AE{}ther coupling $\beta(\phi)$ and the non-canonical kinetic parameter $K_3$.

To obtain the slow-roll limit, we impose that the kinetic contributions are negligible compared to the scalar potential, such that  $H^2 \phi'^2\ll V, H^4 \phi'^4 \ll V$ and $\frac{H'}{H}\ll1$ as the quasi de-sitter solution. Thus Eq.(\ref{eq8}) reduce to

\begin{equation}\label{eq11}
    H^2 =\frac{1}{3\beta}V(\phi)
\end{equation}

Using Eq. \eqref{eq11}, the slow roll condition can be written as
\begin{equation}\label{eq12}
    \phi'^2\ll\beta
\end{equation}

We further impose the condition $H'/H\ll1$ as the quasi-de Sitter condition for a flat expansion rate. Then \eqref{eq9} yields
\begin{equation}\label{eq13}
    \beta'\ll\beta
\end{equation}
so that the effective Lorentz-violating coupling varies slowly throughout the inflationary phase. We further impose the standard slow-roll condition
\begin{equation}\label{eq14}
    \phi''\ll\phi'
\end{equation}
and combining this condition with Eq.\eqref{eq11}, the full equation of motion Eq.\eqref{eq10} reduces to
\begin{equation}\label{eq15}
    \phi'+\frac{V,_\phi}{3H^2}+{\beta,_\phi}=0
\end{equation}
Besides the standard potential gradient term, Eq. \eqref{eq15} also contains an additional term proportional to $\beta_{,\phi}$ signifying the direct effect of Lorentz-violating coupling on the inflation dynamics. In the limit $\beta\to const.$, the extra term disappears, recovering the dynamics of the standard chaotic inflation. Thus, the presence of this term represents a key modification that distinguishes the dynamics of the present model from the standard chaotic inflation.

For the quadratic model \eqref{model1}, we get by solving Eqs. \eqref{eq11} and \eqref{eq15} as
\begin{equation}\label{eq16}
    \phi(\alpha)=\phi_i e^{-4\xi\alpha}
\end{equation}

In this case \eqref{model1}, the Hubble parameter becomes constant
\begin{equation}\label{eq17}
    H^2=\frac{m ^2}{6\xi}
\end{equation}

Since $\phi_i>\phi_c$ and $\phi'^2<\beta$, the condition of $\xi$ becomes
$1/226<\xi<1/16$.

For the quartic model \eqref{model2}, we get by solving Eqs. \eqref{eq11} and \eqref{eq15}
\begin{equation}
    \phi(\alpha)=\frac{\phi_i}{(1+12\phi_i^2\xi\alpha)^{\frac{1}{2}}}
\end{equation}
with the Hubble parameter that is no longer a constant but depends on the field
\begin{equation}
    H^2=\frac{m^2}{6\xi\phi^2}
\end{equation}
The corresponding allowed range of $\xi$ in this model is found to be $1/648\pi<\xi<1/324$. An important difference between the two models is that the quadratic coupling yields a constant Hubble parameter, whereas for quartic coupling, the Hubble parameter evolves with the inflaton field. This distinction will have significant effects on the tensor power spectrum.

\subsection{Standard slow roll regime} \label{GR like slow roll subsection}
Once the inflaton field crosses the critical field value $\phi_c$ ($\phi<\phi_c$), the Lorentz-violating effects become negligible, and the background dynamics is governed by the potential $V(\phi)$. In the standard slow roll regime $8\pi G\beta \ll1$, we have
\begin{equation}
    H^2=\frac{8\pi G}{3} \left(\frac{1}{2} H^2 \phi '^2 +\frac{3K_3}{4} H^4 \phi '^4 +V(\phi)\right)
\end{equation}
\begin{equation}
    \frac{H'}{H}+8\pi G\left(\frac12\phi'^2+K_3H^2\phi'^4\right)=0
\end{equation}
\begin{equation}
\begin{aligned}
       \left(1+3 K_3 H^2 \phi '^2\right)\phi '' &+\left(3+\frac{H'}{H}\right)\phi'\\
       & +4\left(\frac{H'}{H}+1\right) K_3 H^2 \phi'^3
        +\frac{V,_\phi}{H^2}=0 
\end{aligned}  
\end{equation}

The following derivations follow the usual slow-roll analysis, leading to the equations

\begin{equation}\label{eq18}
    H^2=\frac{8\pi G}{3}V
\end{equation}
\begin{equation}\label{eq19}
    \phi'+\frac{V,_\phi}{3H^2}=0
\end{equation}
which are independent of the coupling $\beta(\phi)$, indicating that the Lorentz-violating effects become negligible once the field crosses the critical value

For the quadratic potential $V=\frac{1}{2}m^2\phi^2$, the evolution of the inflaton can be solved as
\begin{equation}\label{eq20}
    \phi^2(\alpha)=\phi_c^2-\frac{\alpha}{2\pi G}
\end{equation}

The scale factor $a(t)=e^{\alpha}$ can also be obtained as
\begin{equation}\label{eq21}
    a(t)=exp[2\pi G(\phi_c^2-\phi^2(t))]
\end{equation}

Inflation ends once the slow-roll condition is violated and the universe enters the reheating phase.

\subsection{e-folding number}
Because the background dynamics have two distinct phases, characterised by $8\pi G\beta(\phi)\gg1$ and $8\pi G\beta(\phi) \ll1$ described in Sec. \ref{LV violating subsection} and Sec. \ref{GR like slow roll subsection} respectively, the total inflationary expansion is the sum of the e-folding numbers in each regime. Let the scalar field value be $\phi_i$ for the e-folding number $N=60$. The total e-folding number for the quadratic model \eqref{model1} is

\begin{equation}\label{efold for model1}
    N=\frac{1}{4\xi}log \frac{\phi_i}{\phi_c}+2\pi G(\phi_c^2-\phi_e^2)
\end{equation}
and for quartic model \eqref{model2}
\begin{equation}\label{efold for model2}
    N=\frac{1}{12\xi}\left(\frac{1}{\phi_i^2}-\frac{1}{\phi_c^2}\right)+2\pi G(\phi_c^2-\phi_e^2)
\end{equation}
where $\phi_e\approx0.3M_{pl}$ is the value of the scalar field at the end of inflation. Note that the first term in Eqs. \eqref{efold for model1} and \eqref{efold for model2} comes from the Lorentz-violating stage.

\section {Potential Free Limit and Analytic Background Solution}
\subsection{Exact solutions}
Before investigating the observational consequences of the tensor sector, we consider a potential-free limit of the background equations. This limit is introduced not as a replacement for the conventional potential-driven inflationary scenario, but as an analytically tractable benchmark that allows us to isolate the effects of the non-canonical kinetic term and the Lorentz-violating gravitational sector. 

In the presence of a non-zero potential, the scalar field and background equations contain additional model-dependent contributions associated with the choice of $V(\phi)$. Setting $V(\phi)=0$ removes this additional dependence and permits the coupled background equations to be solved analytically. The resulting solution provides a direct relation between the scalar field and the scale factor, which can be subsequently be used to express the tensor-sector coupling and the propagation speed of primordial gravitational waves in terms of the inflationary background. 

Setting $V(\phi)=0$ and focusing on the dynamics where non-canonical kinetics dominate ($K_2X\ll K_3X^2$), the background equations \eqref{eq8}-\eqref{eq10}  reduce to, 
\begin{equation}\label{eq22}
    1=\frac{1}{4\beta}K_3H^2\phi'^4
\end{equation}
\begin{equation}\label{eq23}
    \frac{\beta '}{\beta}+\frac{H'}{H}+\frac1\beta\left(K_3H^2\phi'^4\right)=0
\end{equation}
\begin{equation}\label{eq24}
\begin{aligned}
       \left(1+3 K_3 H^2 \phi '^2\right)\phi '' &+\left(3+\frac{H'}{H}\right)\phi'\\ &+4\left(\frac{H'}{H}+1\right) K_3 H^2 \phi'^3
        +3\beta,_\phi =0  
\end{aligned}
\end{equation}

The resulting solution is particularly useful because the scalar field evolution can be expressed directly in terms of the number of e-folds. Consequently, the field dependent tensor coupling can be evaluated at the horizon crossing without requiring a numerical solution of the full background system. Integrating Eq.\eqref{eq23}:
\begin{equation}\label{eq25}
    \beta  H=C e^{-4\alpha} 
\end{equation}
where C is a constant.

The key difference when removing the potential comes from the Friedmann equation \eqref{eq22} and the scalar field equation \eqref{eq24}, as now the dynamics is independent of the scalar potential $V(\phi)$ and solely driven by the Lorentz-violating coupling $\beta(\phi)$ and non-canonical kinetics $K_3$. Substituting Eq. \eqref{eq25} into Eq. \eqref{eq22} to eliminate $H$ in terms of $C$, we obtain an expression of $\phi'$ in terms of $\beta$
\begin{equation}\label{eq26}
    \phi'=-A\beta^{\frac34}e^\alpha
\end{equation}
where we have defined $A=\left(\frac{4}{K_3C^2}\right)^\frac14$. 

The condition for the accelerating universe $\ddot{a} >0$ is now 
\begin{equation}\label{eq27}
    \frac{H'}{H}>-1
\end{equation}
Using Eq. \eqref{eq23} in \eqref{eq27}, we get
\begin{equation}\label{eq28}
    (\log\beta)' < 1
\end{equation}
With Eq. \eqref{eq26}, we obtain the condition on $\beta$ as 
\begin{equation}
    \frac{1}{\beta^{\frac14}}\frac{d\beta}{d\phi}>\frac{3e^{-\alpha}}{A}
\end{equation}

\subsection{\texorpdfstring{$\beta=\xi\phi^2$ model}{beta=xi phi2 model}}
We first consider the quadratic coupling model within the zero potential framework. We obtain the asymptotic form of the solution in the limit $e^{\alpha}\gg1$ near the end of inflation in the Lorentz-violating non-canonical phase. In this limit, the corresponding solution is obtained by integrating Eq. \eqref{eq26}

\begin{equation}\label{eq33}
    \phi=\frac{4}{A^2\xi^{3/2}}e^{-2\alpha}
\end{equation}

This shows that the field decays exponentially with the number of e-folds. Substituting this into Eq.\eqref{eq25} and solving for the Hubble rate gives

\begin{equation}\label{H~exp(2alpha)}
    H=\frac{CA^4}{16}\xi^2e^{2\alpha}
\end{equation}

Using $H=\dot{\alpha}$, Eq. \eqref{H~exp(2alpha)} can be integrated to obtain the scale factor as

\begin{equation}\label{a(t) for phi2}
    a(t)=a_0(t_e-t)^{-\frac{1}{2}}
\end{equation}

where $a_0=(\frac{8}{CA^4\xi^2})^\frac12=2\sqrt{\frac{2CK_3}{\xi^2}}$. This solution describes a super-inflationary regime in which $a(t)$ diverges at finite time $t_e$. This shows that even without a potential, the Lorentz-violating coupling parameter $\xi$ and non-canonical kinetics can drive accelerated expansion.

\subsection{\texorpdfstring{$\beta=\xi\phi^4$ model}{beta=xi phi4 model}}
We next repeat the same analysis for the quartic coupling model. Using the same asymptotic limit $e^\alpha\gg1$ near the end of inflation, the solution is obtained by integrating Eq. \eqref{eq26}

\begin{equation}
    \phi=(2A\xi^{3/4})^{-1/2}e^{-\alpha/2}
\end{equation}

so that, as in the quadratic coupling model, the field decays exponentially with $\alpha$. Substituting this solution into Eq. \eqref{eq25}, the Hubble parameter is found to be

\begin{equation}\label{H=constant}
    H=8\sqrt{\frac{\xi}{K_3}}
\end{equation}

which remains constant throughout this regime. Using $H=\dot{\alpha}$, Eq. \eqref{H=constant} can be integrated to obtain the scale factor as

\begin{equation}\label{a(t) in phi^4}
    a(t)=e^{Ht}
\end{equation}

This solution describes a pure de Sitter expansion whose exact behaviour is governed by the Lorentz-violating coupling parameter and non-canonical kinetics.

\section{Tensor perturbations}
The tensor sector provides a direct probe of the Lorentz violating gravitational dynamics because the aether modifies the kinetic structure and propagation of gravitational waves. Unlike the background equations, which depend on $\beta_1+2\beta_2+\beta_3$, the quadratic tensor action depends on the distinct combination $\beta_1+\beta_3$. This distinction allows primordial gravitational waves to probe a complementary direction in the parameter space of the theory.

Having derived the background solutions, we, therefore, next examine the tensor perturbations about these solutions and their implications. Although the background evolution determines the inflationary universe, the study of perturbation spectra can help test the model against the observational signatures. As the presence of a Lorentz-violating vector field $u^{\mu}$ introduces a preferred frame, the propagation of gravitational waves need no longer coincide with the speed of light. Thus, the field-dependent coupling $\beta_i(\phi)$ modifies the tensor sector and can leave a distinctive imprint on the primordial power spectrum. We now, therefore, focus on the tensor sector, deriving the tensor power spectrum and investigating how the quadratic and quartic models modify the tensor spectral index and its running relative to the standard inflationary prediction.

We work with the full action of Eq. \eqref{action}, reproduced here for convenience

\begin{equation}
\begin{aligned}
    S = \int d^{4} x \sqrt{-g} \bigg[ & \frac{R}{16 \pi G} - \beta_{1}(\phi) \nabla^{\mu} u^{\nu} \nabla_{\mu} u_{\nu} \\
    & - \beta_{2}(\phi) \nabla^{\mu} u^{\nu} \nabla_{\nu} u_{\mu} \\
    & - \beta_{3}(\phi) (\nabla_{\mu} u^{\mu})^{2} \\
    & - \beta_{4}(\phi) u^{\mu} u^{\nu} \nabla_{\mu} u^{\alpha} \nabla_{\nu} u_{\alpha} \\
    & + \lambda (u^{\mu} u_{\mu} + 1) - P(\dot{\phi}, \phi) \bigg]
\end{aligned}
\end{equation}

and perturb the metric, introducing a transverse-traceless tensor $h_{ij}$
\begin{equation}
    ds^{2} = -dt^{2} + a^{2}(t) \left( \delta_{ij} + h_{ij}(t, x^{i}) \right) dx^{i} dx^{j}
\end{equation}

where the perturbations satisfy $h^i_j=h_{ij}^{,j}=0$

Expanding the  action to second-order gives
\begin{equation}\label{2nd action}
    S= \int d^4 x \frac{a^3}{16\pi G}\left[\frac{1}{4}\gamma \dot{h_{ij}} \dot{h^{ij}}-\frac{1}{4a^2}h_{ij,k}h^{ij,k}\right]
\end{equation}

where $\gamma$   is defined by
\begin{equation}\label{gamma}
    \gamma=1-16\pi G(\beta_1+\beta_3)
\end{equation}
is the effective tensor kinetic coefficient. At the quadratic order, the contribution of $\beta_2$ and $\beta_4$ vanishes, and only the contribution of $\beta_1$ and $\beta_3$ survives. The corresponding equation of motion for a Fourier mode $h_k$ follows from varying Eq. \eqref{2nd action}

\begin{equation}\label{PEOM2}
    h''_{k}+\left(2\mathcal{H}+\frac{\gamma'}{\gamma}\right){h'_{k}}+\frac{k^2}{\gamma}h_{k}=0
\end{equation}

The parameter $\gamma$ is dimensionless and represents the propagation speed of tensor perturbations, given by 
\[c_T=\frac{1}{\sqrt{\gamma}}\]

If the parameters $\beta_1=\beta_3=0$, then $\gamma=1$, which is the standard general relativity case. However, if there is a violation of Lorentz symmetry due to the time-like vector field $u^{\mu}$, the tensor kinetic term is modified as shown in Eq.(\ref{2nd action}). As $\beta$ is a function of the inflaton field, the parameter $\gamma$ generally evolves during inflation. Thus, the time-dependent $\gamma$ introduces an extra term $\gamma'/\gamma$ in the tensor-mode equation (\ref{PEOM2}).

Assuming 
\begin{equation}
    h=\frac{u}{z}
\end{equation}
where $z=a\sqrt{\gamma}$, we obtain
\begin{equation}\label{diffeqn}
    u''+\left(\frac{k^2}{\gamma}-\frac{z''}{z}\right)u=0
\end{equation}

Having established the tensor mode equation, we now solve for our two coupling models introduced in Eqs. \eqref{model1} and \eqref{model2}. We first consider the constant $\gamma$ limit as a reference case. This isolates the effect of the background expansion on the tensor spectrum. We subsequently allow $\gamma$ to evolve with the inflaton, thereby capturing the additional contribution associated with the time-dependence of the Lorentz-violating tensor kinetic coefficient. The comparison between these two cases allows us to disentangle background-driven and propagation-driven modifications of the tensor spectrum.

\subsection{\texorpdfstring{Constant $\gamma$}{Constant gamma}}
\subsubsection{\texorpdfstring{For $\beta=\xi \phi^2$ case}{For beta=xi phi\string^2 case}}

For constant $\gamma$, the background solution Eq.\eqref{a(t) for phi2} implies
\[a(t)\propto (t_e-t)^{\frac{- 1}{2}}\] 
Expressed in conformal time, this corresponds to
\[(t_e-t)\propto(-\tau)^{\frac{-1}{3}}\]
so that
\begin{equation}
    a(\tau)=A_0(-\tau)^{\frac{-1}{3}}
\end{equation}

where $A_0=\left(\frac{2a_0^2}{3}\right)^{1/3}=\left(\frac{16CK_3}{3\xi^2}\right)^{1/3}$
\\
Denoting
\begin{equation}\label{vk}
    h_k=\frac{\sqrt{16\pi G }u_k}{z}
\end{equation}

where $z=a\sqrt{\gamma}\propto(-\tau)^{-1/3}$, Eq.\eqref{diffeqn} reduces to a Bessel-type equation
\begin{equation}\label{vkeqn}
    u_k''+\left(\frac{k^2}{\gamma}-\frac{4}{9\tau^2}\right)u_k=0
\end{equation}
whose normalised solution is a Hankel function
\begin{equation}\label{uk bessel eqn}
    u_k=\frac{\sqrt{-\pi \tau}}{2}H_\mu ^{(1)}(x), \quad x=\frac{k(-\tau)}{\sqrt{\gamma}}
\end{equation}
where $\mu=\frac{5}{6}$.

In the super horizon limit, ($x\ll1$), we have $H_\mu ^{(1)}(x)=\frac{i\Gamma(\mu)}{\pi}\left(\frac{x}{2}\right)^{-\mu}$. So, the solution of Eq.\eqref{vkeqn}
\begin{equation}\label{uk approx for const}
    u_k =\frac{i2^{\mu-1}}{\sqrt{\pi}}\Gamma(\mu)(-\tau)^{\frac12-\mu}(c_Tk)^{-\mu}
\end{equation}
Therefore,
\begin{equation}\label{hk with A0}
    h_k=\frac{i2^{\mu-1}}{\sqrt{\pi}}\frac{\Gamma(\mu)}{A_0\sqrt{\gamma}}\sqrt{16\pi G}(c_Tk)^{-\mu}
\end{equation}

with $\mu=5/6$, the conformal time $\tau$ vanishes. Evaluating $A_0$ at the horizon crossing
\begin{equation}
    A_0=\frac{H_k3^{1/3}}{(c_Tk)^{4/3}}
\end{equation}
and substituting it in Eq.\eqref{hk with A0} gives the mode amplitude
\begin{equation}
    |h_k|^2=2^{2\mu+2}G\Gamma^2(\mu)3^\frac23\frac{H_k^2}{c_Tk^3}
\end{equation}
for which the tensor power spectrum of gravitational waves follows as
\begin{equation}\label{PT for phi2}
\begin{aligned}
   & P_T=2\times\frac{2k^3}{\pi^3}|h_k|^2 \\
   & P_T=\frac{4}{\pi^3}2^{2\mu+2}G\Gamma^2(\mu)3^\frac23\frac{H_k^2}{c_T}
\end{aligned}
\end{equation}

At horizon crossing ($k=a_kH_k$),
\[k\propto (t_e-t)^{-\frac32}\]

Since $H(t)=\frac{\dot{a}}{a}\propto \frac1t$ and $a(t)\propto {t^{-\frac12}}$, the Hubble rate at horizon crossing scales as
\[H_k\propto k^{\frac23}\]

and the tensor spectral index can be found as 
\begin{equation}\label{nT=4/3}
\begin{aligned}
  &  n_T=\frac{dlnP_T}{dlnk}\\
  &  n_T=2\frac{dlnH_k}{dlnk}\\
  &  n_T=\frac43
\end{aligned}
\end{equation}
indicating a strongly blue-tilted tensor spectrum.

\subsubsection{\texorpdfstring{For model $\beta=\xi\phi^4$}{For model beta = xi phi\string^4}}
For the quartic model, the constant Hubble parameter solution in Eq. \eqref{a(t) in phi^4} in conformal time ($\tau$) gives

\begin{equation}\label{scale factor phi^4 in conformal time}
    a(\tau)=\frac{1}{H(-\tau)}
\end{equation}
\\
Substituting Eq.\eqref{scale factor phi^4 in conformal time} into Eq. \eqref{diffeqn} and using  Eq.\eqref{vk}, we get
\begin{equation}
    u_k''+\left(\frac{k^2}{\gamma}-\frac{2}{\tau^2}\right)u_k=0
\end{equation}
\\
Solving similarly to equation \eqref{uk bessel eqn}, we get
\begin{equation}
    u_k=\frac{\sqrt{-\pi \tau}}{2}H_\mu ^{(1)}(x), \quad x=\frac{k(-\tau)}{\sqrt{\gamma}}
\end{equation}
where \[\mu=\frac32\]
\\
Solving for the super horizon case ($x\ll1$), we get
\begin{equation}
    u_k=\frac{-i}{\sqrt{2}}k^{-\frac{3}{2}}(-\tau)^{-1}\gamma^{-\frac{3}{4}}
\end{equation}

Therefore,
\begin{equation}
    h_k\propto k^{-\frac{3}{2}}\gamma^{\frac14}
\end{equation}

So, for this case, the tensor power spectrum of gravitational waves is given by,
\begin{equation}
\begin{aligned}
&   P_T(k)=2\times \frac{2k^3}{\pi^2}|h_k|^2\\
&   P_T(k)=\frac{32G}{\pi} H^2 \gamma^{\frac12}\\
&   P_T(k) = constant
\end{aligned}   
\end{equation}

As $H$ is a constant here, $P_T$ doesn't have explicit $k$-dependence at fixed $\gamma$ and the tensor spectral index is exactly invariant,
\begin{equation}\label{nT=0}
    n_T=0
\end{equation}

The constant $\gamma$ approximation shows that both coupling models show distinct tensor spectra reflecting the different background dynamics of each model. The quadratic coupling model $\beta=\xi\phi^2$ yields a strongly blue-tilted spectrum, which reflects the underlying super-inflationary background evolution of the universe. In contrast, the quartic model $\beta=\xi\phi^4$ remains exactly scale-invariant, which clearly reflects the de-Sitter background evolution. Physically, the increasing Hubble rate for the quadratic model causes modes with larger $k$ to exit the horizon at later times when $H$ is larger, increasing power and hence a blue-tilted spectrum. In the de Sitter background of the quartic model, as $H$ remains constant at horizon crossing, the tensor spectrum remains scale-invariant.

\subsection{\texorpdfstring{Time dependent $\gamma$}{Time dependent gamma}}\label{Sec time dependent gamma}
\subsubsection{\texorpdfstring{For $\beta=\xi\phi^2$}{For beta = xi phi\string^2}}\label{subsection phi2 time dependent}
We next consider the general case, where $\gamma$ evolves during inflation. In the super horizon limit($k\ll\mathcal{H}$) on Eq.\eqref{diffeqn} reduces to

\begin{equation}
    u''-\frac{z''}{z}u=0
\end{equation}
The solution of the above equation is given by
\begin{equation}\label{solnofdifferential}
    u(\tau)=C_1(k)z+C_2(k)z \int_{0}^{\tau} \frac{d\tau'}{z^2}
\end{equation}

The first term of the solution is the dominant term.\\
For deep subhorizon, we have
\[c_Tk\gg aH\]\\
then,
\begin{equation}
    u_k=\frac{e^{-ic_Tk\tau}}{\sqrt{2c_Tk}}
\end{equation}

Matching with the superhorizon, we get
\begin{equation}\label{solution of uk after bunch davies}
    u_k(\tau)\sim\frac{1}{\sqrt{2c_Tk}}
\end{equation}
So, from Eq. (\ref{solnofdifferential}), taking the dominant term $(u(\tau)\approx C_1(k)z)$ and applying the horizon crossing condition ($a_k=\frac{c_Tk}{H_k}$), we get
\begin{equation}
    C_1(k)\sim\frac{H_k}{\sqrt{2}}\gamma_k^{\frac{1}{4}}k^{\frac{-3}{2}}
\end{equation}

The tensor power spectrum of a gravitational wave is given by
\begin{equation}\label{P S for varying g}
    P_T=\frac{k^3}{2\pi}\left|h_k\right|^2
\end{equation}
As $h_k=\sqrt{16\pi G}C_1(k)$
So, we get from Eq.\eqref{P S for varying g}
\begin{equation}
    P_T=\frac{32}{\pi} GH_k^2\gamma^{\frac12}
\end{equation}

We make an approximation

\begin{equation}\label{gamma=1-q}
\begin{aligned}
&\gamma=1-16\pi G(\beta_1+\beta_3)\\
&\gamma   \sim 1-\lambda_T \phi^2 \\
&\gamma   =1-q
\end{aligned}
\end{equation}
where $q=\lambda_T\phi^2\propto a(t)^{-4}$ and $\lambda_T\sim 16\pi G f(\xi)$ as $\beta_1+\beta_3=f(\xi)\phi^n$, $n=2, 4$

Since $P_T\propto H_k^2\sqrt{\gamma_k}$ and $\gamma$ is time-dependent, we get by solving for $n_T$
\begin{equation}\label{nT for phi2 general}
\begin{aligned}
  &  n_T=\frac{dlnP_T}{dlnk}\\
  &  n_T=\frac{dlnP_T}{d\alpha}\times \frac{dln\alpha}{dlnk} \\
  &  n_T= \frac{4-2q}{3-q}
\end{aligned}
\end{equation}

For $q\ll1$,
\begin{equation}\label{nT for phi2}
    n_T=\frac43-\frac29 q
\end{equation}
where $\frac29 q$ is the small correction term proportional to $q$. Now, the corresponding running of the tensor spectral index is given by
\[\alpha_T=\frac{dn_T}{dlnk}\]
Therefore, by solving, we get
\begin{equation}\label{running of spectral index}
    \alpha_T=\frac{8q(1-q)}{(3-q)^3}
\end{equation}

For $q\ll1$, 
\begin{equation}\label{alphaT for phi2}
    \alpha_T\sim \frac{8}{27}q
\end{equation}
which is a positive quantity. Eqs. \eqref{nT for phi2}and \eqref{alphaT for phi2} are two model-specific relations to place observational bounds on the coupling parameter.

\subsubsection{\texorpdfstring{For $\beta=\xi\phi^4$}{For beta = xi phi\string^4}}\label{subsection phi4 time dependent}
In this case, we have from  Eq. \eqref{scale factor phi^4 in conformal time},
\[a(\tau)=\frac{1}{H(-\tau)}\]
\\
Following the $\phi^2$ case, we obtain a similar solution as in Eqs.\eqref{solnofdifferential} and \eqref{solution of uk after bunch davies}. So, after applying the horizon crossing condition, we get
\begin{equation}
     C_1(k)\sim\frac{H_k}{\sqrt{2}}\gamma_k^{\frac{1}{4}}k^{\frac{-3}{2}}
\end{equation}
So, the tensor power spectrum in this case is found to be
\begin{equation}
    P_T=constant\times \gamma^{\frac12}
\end{equation}
\\
The tensor spectral index, therefore, is 
\begin{equation}
    n_T=\frac12 \frac{dln \gamma_k}{d lnk}
\end{equation}
\\
Here
\begin{equation}
\begin{aligned}
\gamma &  \sim 1-\lambda_T \phi^4 \\
&   =1-q
\end{aligned}
\end{equation}
where $q=\lambda_T\phi^4\propto a(t)^{-2}$

Since $H$ is constant, at horizon crossing $c_Tk=aH$, we have
\[P_T\propto \sqrt{\gamma}\]

Therefore,
\begin{equation}\label{nT for phi4 general}
    n_T=q
\end{equation}

For $q\ll1$.
\begin{equation}\label{nT for phi4}
    n_T\approx q
\end{equation}

And its running is given by
\begin{equation}\label{alphaT for phi4 general}
    \alpha_T=-2q(1-q)
\end{equation}

In the limit, $q\ll1$,
\begin{equation}\label{alphaT for phi4}
    \alpha_T\approx-2q
\end{equation}

The evolution of $\gamma$ introduces distinct corrections to the tensor sector in the two models. For the quadratic coupling model $\beta=\xi\phi^2$, there is a shift in the tensor spectral index from its constant $\gamma$ value by a small negative correction, with a corresponding positive running. In contrast, for the quartic coupling model $\beta=\xi\phi^4$, the tensor spectrum acquires a small positive tilt and a negative running.

\section{Consistency relations}
A remarkable aspect of inflationary theory is the existence of consistency relations between different cosmological observable signatures. These relations provide a direct test of the underlying theoretical framework for the universe's inflationary dynamics. These relations are significant as they impose stringent bounds on theoretical models through observational data. In our present Lorentz-violating framework with non-canonical kinetics, the tensor sector is modified through the field-dependent tensor propagation factor $\gamma(\phi)$. As a result, the tensor spectral index and its running are no longer independent quantities, but satisfy model-dependent consistency relations. 

These relations arise as both observables depend on the effective Lorentz-violating coupling parameter given by Eq. \eqref{gamma=1-q}. Eliminating the parameter $q$ from the relations of $n_T$ and $\alpha_T$ yields a consistency relation solely in terms of the observable quantities. Such relations provide robust predictions to our theoretical framework.

\subsection{\texorpdfstring{The $\beta=\xi\phi^2$ model}{The beta = xi phi\string^2 model}}
The consistency relation for the quadratic coupling model is obtained from Eqs. \eqref{nT for phi2 general} and \eqref{running of spectral index}. Eliminating $q$, we get
\begin{equation}\label{nT and alphaT for phi2 general}
    \alpha_T=2(3n_T-4)(n_T-1)(n_T-2)
\end{equation}
and the approximate expression in the limit $q\ll1$ is obtained from Eqs. \eqref{nT for phi2} and \eqref{alphaT for phi2}

\begin{equation}\label{nT and alphaT for phi2}
    12n_T=16-9\alpha_T+\mathcal{O}(q^2)
\end{equation}

Eqs.\eqref{nT and alphaT for phi2 general} and \eqref{nT and alphaT for phi2} provide a distinctive prediction of our inflationary model. This establishes a unique relation between $n_T$ and $\alpha_T$, implying that they do not vary independently. This relation strengthens the predictive power of our model and offers a direct observational test of our model. Any future measurements of $n_T$ and $\alpha_T$ consistent with this relation would support the quadratic coupling model.

\subsection{\texorpdfstring{The $\beta=\xi\phi^4$ model}{The beta = xi phi\string^4 model}}
For the quartic coupling model, replacing $q$ in Eq. \eqref{alphaT for phi4 general} with $n_T$ from Eq. \eqref{nT for phi4 general}, we get
\begin{equation}\label{nT and alphaT phi4 general}
    \alpha_T=-2n_T(1-n_T)
\end{equation}

And the approximate expression in the limit $q\ll1$, is given by Eqs.\eqref{nT for phi4} and \eqref{alphaT for phi4}, 

\begin{equation}\label{nT and alphaT for phi4}
    \alpha_T=-2n_T+\mathcal{O}(q^2)
\end{equation}

Eq. \eqref{nT and alphaT for phi4} provides the characteristic prediction of the quartic inflationary model. The linear form of Eq. \eqref{nT and alphaT for phi4} indicates that once the tensor spectral index is specified, its running is uniquely determined. As a result, this reduces the number of independent observables required to characterise the tensor spectrum. The resulting consistency relation thus provides a direct observational test for the quartic coupling model of inflationary dynamics.

\subsection{Tensor stability}
Before discussing observational implications, it is useful to examine the stability of the tensor sector. As the Lorentz-violating coupling modifies the kinetic coefficient of the tensor perturbations through $\gamma$, the stability requires a theoretical constraint on the parameter $q$ during inflation. 

The tensor perturbations remain stable provided
\begin{equation}
    \gamma>0
\end{equation}

For both of the coupling models, we have
\begin{equation}
    \gamma=1-q
\end{equation}
and therefore the stability condition becomes
\begin{equation}
\begin{aligned}
     1-q>0\\
     q<1
\end{aligned}
\end{equation}

\section{Observational Constraints and Forecasts}
In this section, we test our Lorentz-violating non-canonical inflationary model with the current observational data and sensitivities expected from future experiments. We map our model parameters to the observables such as gravitational wave propagation speed $c_T$, tensor spectral index $n_T$ and its running spectral index $\alpha_T$, thereby establishing empirical bounds on parameters.

The multimessenger observation of GW170817\cite{Abbott2017GW170817}  and GRB170817A \cite{Abbott2017GRB170817A} provides stringent constraints on the propagation speed of gravitational waves at low redshifts, requiring
\begin{equation}\label{GW speed constriant}
    \left|\frac{c_T-c}{c}\right|\leq 10^{-15}
\end{equation}
Therefore,
\[\gamma\sim 1\]
Expanding
\[c_T=\frac{1}{\sqrt{\gamma}}\] we get
\begin{equation}
    |\gamma-1|\leq 10^{-15}
\end{equation}
Since
\[\gamma-1=-16\pi G(\beta_1+\beta_3)\]
Therefore,
\begin{equation}
    16\pi G|\beta_1+\beta_3| \leq 10^{-15}
\end{equation}
\\
In evaluating observational bounds, we adopt natural units with reduced Planck mass $M_{pl}=(8\pi G)^{-1/2}=1$

\subsection{\texorpdfstring{For $\beta=\xi \phi^2$}{For beta = xi phi\string^2}}
From Eq.\eqref{PT for phi2}, for almost scale invariant ($\mu=5/6, \Gamma(\frac56)$), we have
\begin{equation}
    P_T= \frac{4.3H^2}{M_{pl}^2 c_T}
\end{equation}
with $c_T=\frac{1}{\sqrt{\gamma}}$
\\
Since 
\[P_T=rP_R\] with $r<r_{max}$, 
\begin{equation}
    H^2\sqrt{\gamma}<\frac{M_{pl}^2}{4.3}r_{max}P_R
\end{equation}
Presently, we have $r <0.03$\\
and, $P_R\approx 2.1\times 10^{-9}$, 
\[H<3.83\times 10^{-6} M_{pl}\]
\\
For $\gamma\approx1$, 
\begin{equation}
    H\leq 9.3 \times 10^{12}GeV
\end{equation}
\\
For $\gamma\not\approx1$
\begin{equation}
    H_{max}\propto \gamma^{-\frac{1}{4}}
\end{equation}

We now use the model-dependent expressions for the tensor spectral index $n_T$ and its running $\alpha_T$, obtained in Sec. \ref{subsection phi2 time dependent}, Eqs.\eqref{nT for phi2} and \eqref{alphaT for phi2}, and derive three separate bounds on the effective coupling $q$, where $\gamma=1-\lambda_T\phi^2=1-q$. The constraints derived below depend on the small correction term at the horizon crossing $q_k$.


\subsubsection{Primordial early time constraints}
\paragraph{Tensor spectral index $n_T$}:
The tensor spectral index $n_T$ directly probes the evolution of $\gamma$. In the constant-$\gamma$ limit, the quadratic coupling model predicts a fixed blue-tilt spectrum with $n_T=4/3$ as shown in Eq.\eqref{nT=4/3}, while the evolution of $\gamma$ introduces the extra correction term given in Eq.\eqref{nT for phi2}, and it is this contribution that we use to constrain $q_k$ at the horizon crossing.

 Future CMB and gravitational wave experiments like LiteBIRD, CMB-S4, DECIGO aim to measure $n_T$ to the level of $10^{-2}-10^{-3}$ \cite{LiteBIRD:2023ptep,LiteBIRD:2024jcap,CMBS4:2016book,CMBS4:2017tech,DECIGO:2011cqg,DECIGO:2021ptep}. Therefore, the correction term in Eq.\eqref{nT for phi2} has to remain below this level to predict its significance.

\begin{equation}\label{q inequality for nT in phi2}
\begin{aligned}
  &  |\nabla n_T|\leq 10^{-2}\\
  &  \frac{2}{9}|q_k|\lesssim 10^{-2}
\end{aligned}
\end{equation}

Therefore, the constraint on $q_k$ is
\begin{equation}
    -4.5\times 10^{-2}\lesssim q_k \lesssim 4.5 \times 10^{-2}
\end{equation}
indicating the future measurements of the tensor spectral index that could constrain the Lorentz-violating coupling at horizon crossing.

\paragraph{Running index $\alpha_T$}:
From future observations like LiteBIRD, CMB-S4, DECIGO, and BBO, it is expected to reach sensitivities roughly around $\alpha_T\leq 10^{-4}-10^{-3}$ \cite{LiteBIRD:2023ptep,LiteBIRD:2024jcap, CMBS4:2016book,CMBS4:2017tech,DECIGO:2011cqg,DECIGO:2021ptep,BBO:2006prd,BBO:2005prd,BBO:2006cqg}. 
Suppose 
\[|\alpha_T|\leq10^{-4}\]

then from Eq.\eqref{alphaT for phi2}, we have constraint on the coupling $q$ at the horizon crossing as

\begin{equation}
\begin{aligned}
    & |q_k| \lesssim \frac{27\times 10^{-4}}{8}
\end{aligned}
\end{equation}

which is 
\begin{equation}
    -3.34\times 10^{-4}\lesssim q_k \lesssim 3.34\times 10^{-4}
\end{equation}

\subsubsection{Late time constraint}\label{late time constraint}
To connect the inflationary tensor sector with the present-day gravitational wave observational constraint, we assume that the tensor perturbation speed remains close to its current value throughout cosmological evolution. Thus, the stringent GW170817/GRB170817A \cite{Abbott2017GW170817,Abbott2017GRB170817A} constraint on the tensor perturbation speed $c_T$ can be expressed as a corresponding limit on the coupling $q$ given by Eq.\eqref{GW speed constriant}. This allows late-time gravitational wave observations to test whether the tensor sector remains consistent with present-day gravitational observations.

We solve for the first order in $10^{-15}$ with $c=1$ and  $|c_T-1|\lesssim10^{-15}$. As $c_T=\gamma^{-1/2}$, we have 
\begin{equation}
    -2\times 10^{-15}\lesssim q_{GW}\lesssim 2\times 10^{-15}
    \label{qgwquadratic}
\end{equation}
where $q_{GW}$ denotes the value of the tensor sector coupling at the redshift of GW170817 \cite{Abbott2017GW170817}. It is important to distinguish this late-time constraint from the primordial value $q_k$ evaluated at the horizon crossing. A direct identification $q_k=q_{GW}$ requires the tensor-sector coupling to remain approximately unchanged during the evolution from inflation to the late universe. Under this assumption, the GW170817 bound may be interpreted as a conditional primordial constraint.

\subsection{\texorpdfstring{For $\beta=\xi\phi^4$}{For beta = xi phi\string^4}}

\subsubsection{Primordial early time constraints}
\paragraph{Tensor spectral index $n_T$}:
In contrast to the quadratic model, the quartic coupling model predicts an exactly scale-invariant tensor spectrum in the constant $\gamma$ limit, $n_T=0$ as shown in Eq.\eqref{nT=0}. As a result, any deviation from scale invariance directly measures the Lorentz-violating contribution. As the constant $\gamma$ doesn't have any contribution in the quartic model, the tensor tilt arises entirely from the field dependence of $\gamma$ and is approximately given by $n_T\approx q$

Imposing $|\nabla n_T|\leq10^{-2}$, we have
\begin{equation}\label{q inequality for nT in phi4}
\begin{aligned}
  &  |\nabla n_T|\leq 10^{-2}\\
  &  |q|\lesssim 10^{-2} 
\end{aligned}
\end{equation}

Therefore, the constraint on $q_k$ at horizon crossing
\begin{equation}
    -10^{-2}\lesssim q_k \lesssim 10^{-2}
\end{equation}

\paragraph{Running index $\alpha_T$}:
The running of the tensor spectral index further distinguishes the quartic model from its quadratic model. According to Eq.\eqref{alphaT for phi4}, the running of the spectral index yields a negative value, while the quadratic model predicts a positive value. This would provide a distinctive signature of the quartic coupling scenario. Imposing

\[|\alpha_T|\leq10^{-4}\]

then from Eq.\eqref{alphaT for phi4}, we have

\begin{equation}
\begin{aligned}
    & |q| \lesssim \frac{10^{-4}}{2}
\end{aligned}
\end{equation}

Therefore, the constraint on $q_k$ at the horizon crossing is 
\begin{equation}
    -0.5\times10^{-4}\lesssim q_k \lesssim 0.5\times10^{-4}
\end{equation}

\subsubsection{Late time constraint}
Similar to the quadratic coupling model, an independent constraint on $q_{GW}$ can be obtained from the late-time astrophysical measurements of gravitational wave speed. Following the same procedure as in Sec. \ref{late time constraint} and expanding $c_T=\gamma^{-1/2}$ to the first order in $10^{-15}$ with $c=1$,  $|c_T-1|\lesssim10^{-15}$, we have
\begin{equation}\label{q_GW constraint}
    -2\times 10^{-15}\lesssim q_{GW}\lesssim 2\times 10^{-15}
\end{equation}

\subsection{\texorpdfstring{Evolution of gravitational wave coupling $q_{\rm GW}$}{Evolution of gravitational wave coupling q\string_GW}}
The constraints derived in Eqs. \eqref{qgwquadratic} \& \eqref{q_GW constraint} apply to the post-inflationary regime, whereas our analysis focuses on the inflationary epoch. To relate the inflationary constraints to the present-day coupling $q_{GW}$, we introduce an evolution parameter that describes the evolution of the gravitational-wave coupling from inflation to the present epoch (assuming that the effective coupling parameter $\lambda_T$ does not evolve significantly). As we know
\[q_k=\lambda_T\phi_k^p  \qquad   q_{GW}=\lambda_T\phi^p\]
with $p=2,4$. The evolution factor $\chi_p$ given by 
\begin{equation}
    \chi_p=\left(\frac{\phi_{GW}}{\phi_k}\right)^p
\end{equation}
then
\begin{equation}
    q_{GW}=\chi_pq_k
\end{equation}

If the quantity controlling the tensor speed evolves slowly, we can take $\chi_p\approx 1$. But if $\chi \neq1$, we can study $\chi_p$ as an evolution parameter.
\[\chi_p=1,10^{-1},10^{-2},10^{-3},...\]
For a given primordial constraint
\begin{equation}
    |q_k|\lesssim q_{k,max}
\end{equation}
and there
\begin{equation}
    |q_{GW}|\lesssim \chi_pq_{k,max}
\end{equation}

Combining this with the GW bound
\begin{equation}
    |q_{GW}|\lesssim q_{GW,max}
\end{equation}
gives
\begin{equation}
    \chi_p \lesssim\frac{q_{GW,max}}{q_{k,max}}
\end{equation}
Now, for the quadratic coupling model ($\beta=\xi\phi^2$), if $|q_k|\lesssim 0.045$, which is the maximum allowed value from primordial early-time constraints, the evolution parameter must be
\begin{equation}
    \chi_2\lesssim 5.6\times 10^{-14}
\end{equation}
while for the quartic coupling model $\beta=\xi\phi^4$, if $|q_k|\lesssim 0.010$, the evolution parameter must be
\begin{equation}
    \chi_4\lesssim 2.5\times 10^{-13}
\end{equation}
Under the assumption that the effective coupling $\lambda_T$ itself does not evolve significantly, the bounds obtained on $\chi_p$ indicate that the tensor-sector coupling $q$ must be strongly suppressed between the horizon crossing and late-time universe.

\onecolumngrid
\vspace{1em}
\begin{center}
\begin{minipage}{0.95\linewidth}
\captionof{table}{Constraints on the tensor-sector coupling obtained from primordial tensor observables at the horizon crossing and late time gravitational wave propagation speed. The primordial constraint apply to $q_k$, whereas the GW170817 constraint applies to $q_{GW}$.  }
\label{tab:q_constraints}
\begin{ruledtabular}
\begin{tabular}{cccccc} 

Model  & $n_T$ constraint on $q_k$ & $\alpha_T$ constraint on $q_k$ & $c_T$ constraint on $q_{GW}$ \\ \hline

$\beta=\xi\phi^2$

& $\pm 4.5\times 10^{-2}$
& $\pm 3.4\times 10^{-4}$
& $\pm 2\times 10^{-15}$\\ \hline

$\beta=\xi\phi^4$

& $\pm 10^{-2}$
& $\pm 0.5\times 10^{-4}$
& $\pm 2\times10^{-15}$ \\

\end{tabular}
\end{ruledtabular}
\end{minipage}
\end{center}
\vspace{1em}
\twocolumngrid

Table \ref{tab:q_constraints} summarises the constraints on the coupling parameter obtained from tensor-sector observable quantities. These provide a direct connection between the observations and the underlying parameter space of the model.
\section{Comparative study of models}
A comparative study of the tensor-sector observables reveals clear phenomenological differences between the quadratic and quartic field-dependent \AE{}ther couplings. These differences originate from the distinct evolution of the effective tensor sector coupling during inflation. For the quadratic and quartic cases, respectively, 
\begin{equation}
    q\propto \phi^2\propto a^{-4} \quad q\propto \phi^4\propto a^{-2}
\end{equation}
Thus, the effective Lorentz-violating contribution decreases more rapidly in the quadratic model, whereas it persists for a longer period in the quartic model. This different evolution is directly reflected in the tensor spectral index, its running, and the evolution of the tensor propagation factor $\gamma$.

\textbf{A. Tensor spectral index and running as a function of $q$:}

\begin{center}
\begin{figure*}[h!]
    \centering
    \begin{minipage}{0.48\textwidth}
        \centering
        \includegraphics[width=\textwidth]{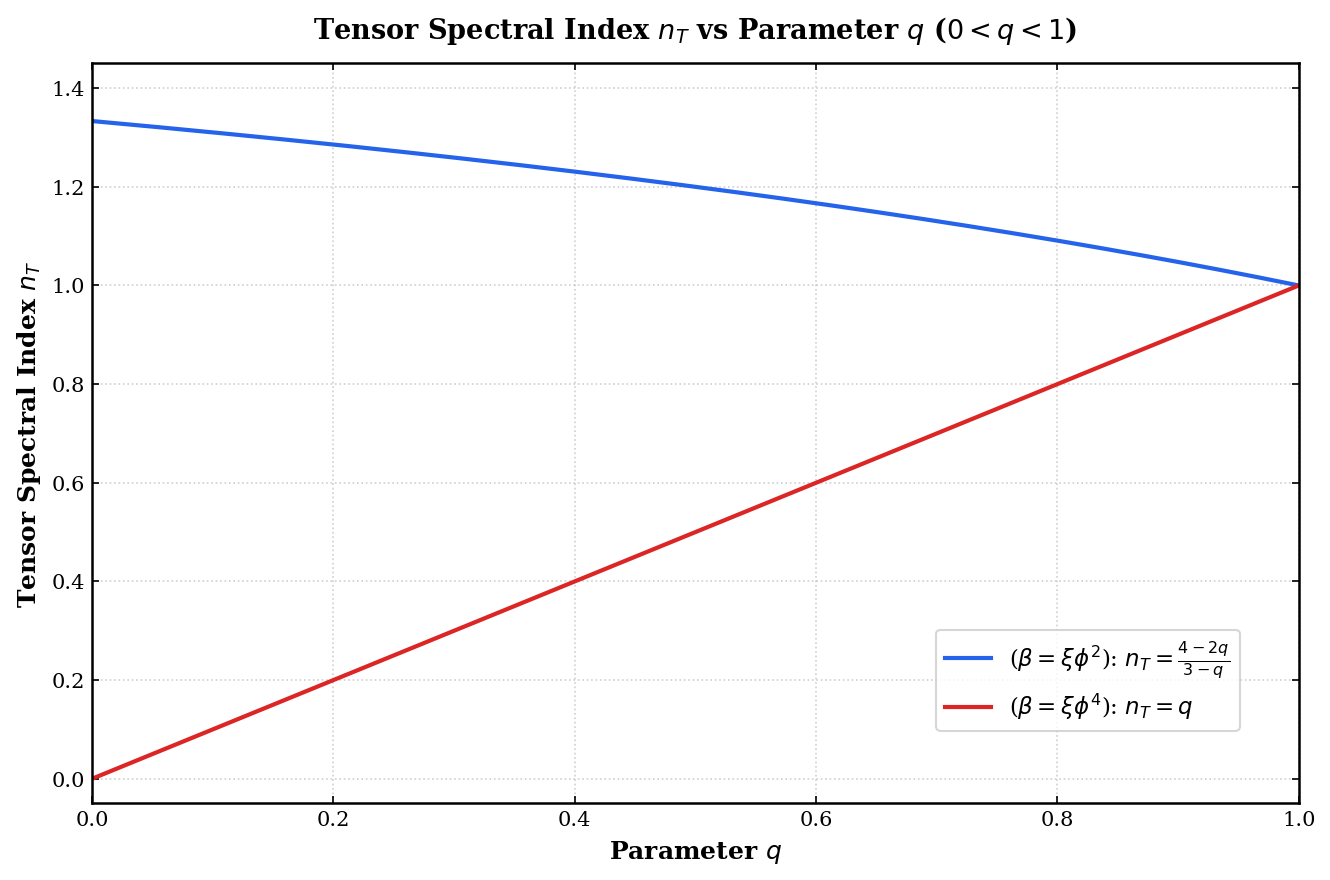}\\
        (a) 
        \label{fig:plot11}
    \end{minipage}\hfill
    \begin{minipage}{0.48\textwidth}
        \centering
        \includegraphics[width=\textwidth]{"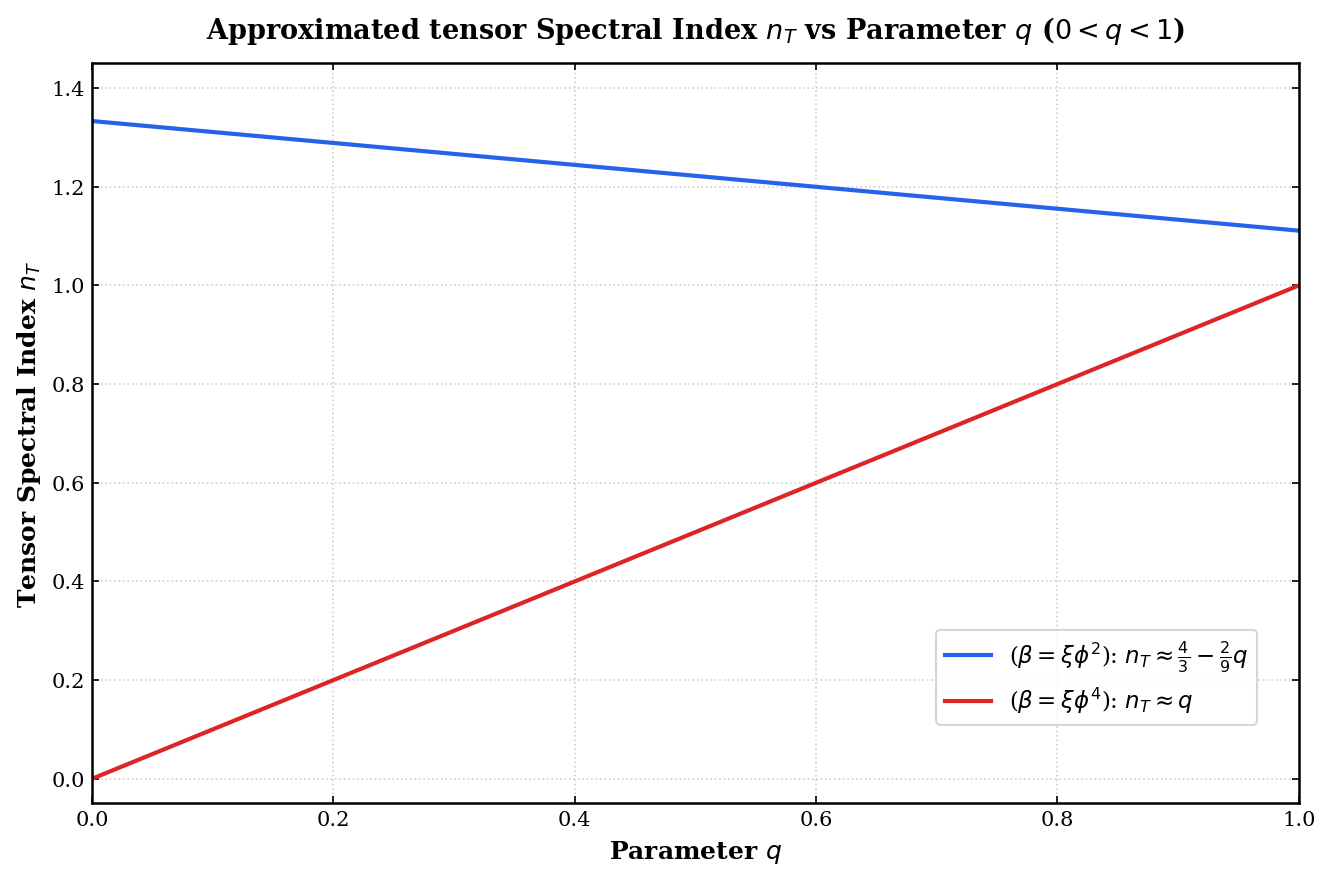"}\\
        (b) 
        \label{fig:plot5}
    \end{minipage}
    \caption{Tensor spectral index $n_T$ as a function of Lorentz violating coupling $q$ for (a) quadratic coupling $\beta=\xi \phi^2$ (b) quartic coupling $\beta=\xi \phi^4$. The exact relations are shown together with their exact model predictions.}
    \label{fig:nT_vs_q}
\end{figure*}
\end{center}

\begin{figure*}
    \centering
    \begin{minipage}{0.48\textwidth}
        \centering
        \includegraphics[width=\textwidth]{"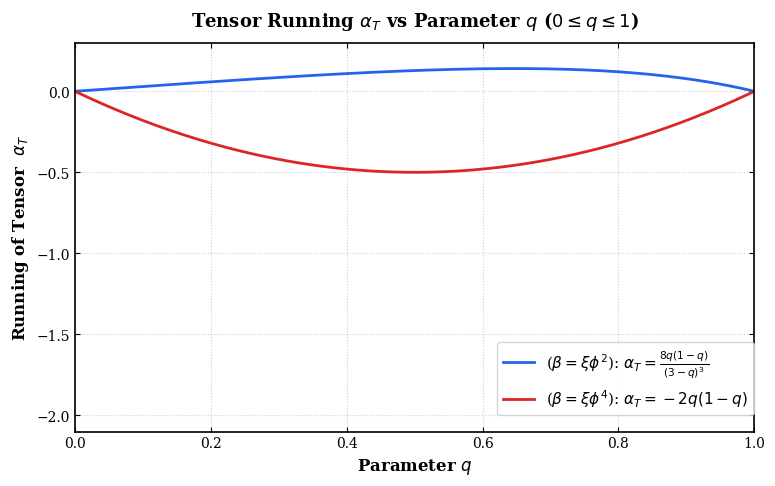"}\\
        (a)
        \label{fig:plot1}
    \end{minipage}\hfill
    \begin{minipage}{0.48\textwidth}
        \centering
        \includegraphics[width=\textwidth]{"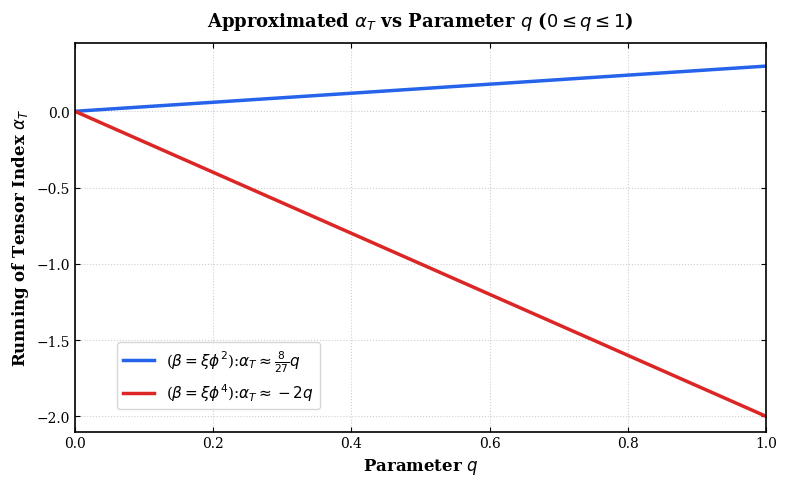"}\\
        (b) 
        \label{fig:plot3}
    \end{minipage}
    \caption{Tensor running $\alpha_T$ as a function of Lorentz violating coupling $q$ for (a) quadratic coupling $\beta=\xi \phi^2$ (b) quartic coupling $\beta=\xi \phi^4$. The exact and small-$q$ behaviour illustrates the distinct running predicted by the two models.}
    \label{fig:alphaT_vs_q}
\end{figure*}

Figs. \ref{fig:nT_vs_q} and \ref{fig:alphaT_vs_q} compare the exact tensor spectral index $n_T$ and its running $\alpha_T$ with the effective coupling parameter $q$ for the two models. Fig. \ref{fig:nT_vs_q} shows that the exact and small-coupling expressions for $n_T$ remain very close over most of the physically relevant range $q<1$, with noticeable differences only as $q$ approaches unity.
Fig. \ref{fig:alphaT_vs_q} shows a more pronounced distinction for the running $\alpha_T$, because the exact result contains higher-order contributions in $q$ that are absent in the $q\ll1$ approximation. This behaviour illustrates that $\alpha_T$ provides a more sensitive diagnostic of departures from the weak Lorentz-violating regime. The small coupling expressions are therefore reliable in the weak coupling regime, while the exact relations should be retained when exploring the full theoretical parameter space.

\begin{figure*}
    \centering
    \begin{minipage}{0.48\textwidth}
        \centering
        \includegraphics[width=\textwidth]{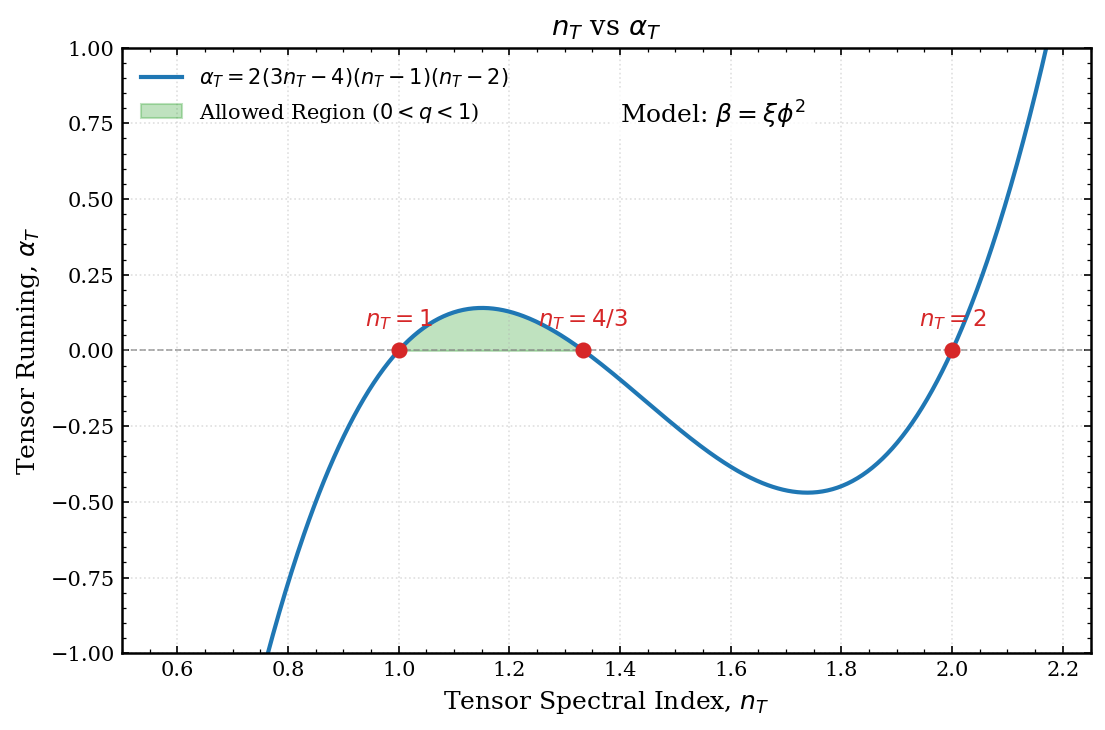}\\
        (a) 
        \label{fig:plot8}
    \end{minipage}\hfill
    \begin{minipage}{0.48\textwidth}
        \centering
        \includegraphics[width=\textwidth]{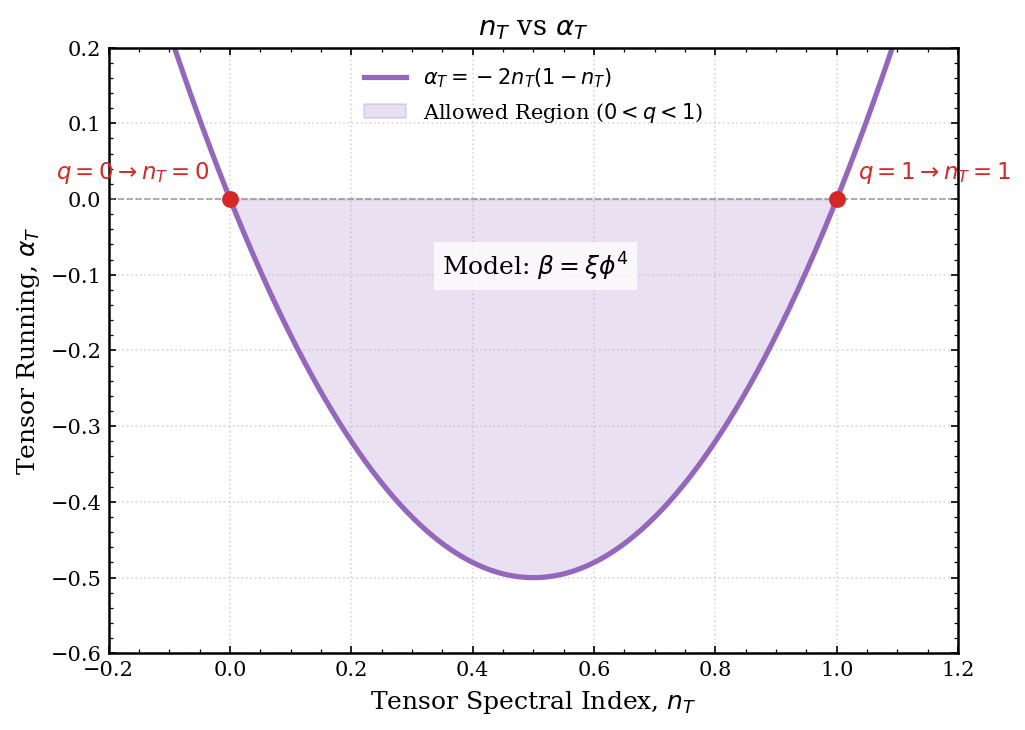}\\
        (b) 
        \label{fig:plot9}
    \end{minipage}
    \caption{Exact consistency relations between the tensor spectral index $n_T$ and its running $\alpha_T$ for the (a) quadratic coupling $\beta=\xi\phi^2$ and (b) quartic coupling $\beta=\xi\phi^4$. The shaded region corresponds to the physically allowed range $0<q<1$.}
    \label{fig:consistency_phi2_phi4}
\end{figure*}

The difference between the two models becomes more significant in the $n_T-\alpha_T$ plane. Fig. \ref{fig:consistency_phi2_phi4} presents the exact consistency relations between the tensor spectral index $n_T$ and its running $\alpha_T$. For the quadratic coupling, eliminating $q$ gives $\alpha_T=2(3n_T-4)(n_T-1)(n_T-2)$, whereas for the quartic coupling one obtains $\alpha_T=-2n_T(1-n_T)$. The two relations have qualitatively different shapes and predict different signs of $\alpha_T$ over their respective physically allowed regions. Thus
\begin{equation}
    \alpha_T>0 \quad {\rm for}\quad \beta=\xi \phi^2
\end{equation}
whereas
\begin{equation}
    \alpha_T<0 \quad {\rm for}\quad \beta=\xi \phi^4
\end{equation}
This opposite sign of the running provides a particularly clean theoretical discriminator between the two functional forms of the Lorentz violating couplings.
A simultaneous measurement of $n_T$ and $\alpha_T$ can therefore provide a direct discriminator between the two coupling scenarios.

\begin{figure*}
    \centering
        \begin{minipage}{0.48\textwidth}
        \centering
        \includegraphics[width=\textwidth]{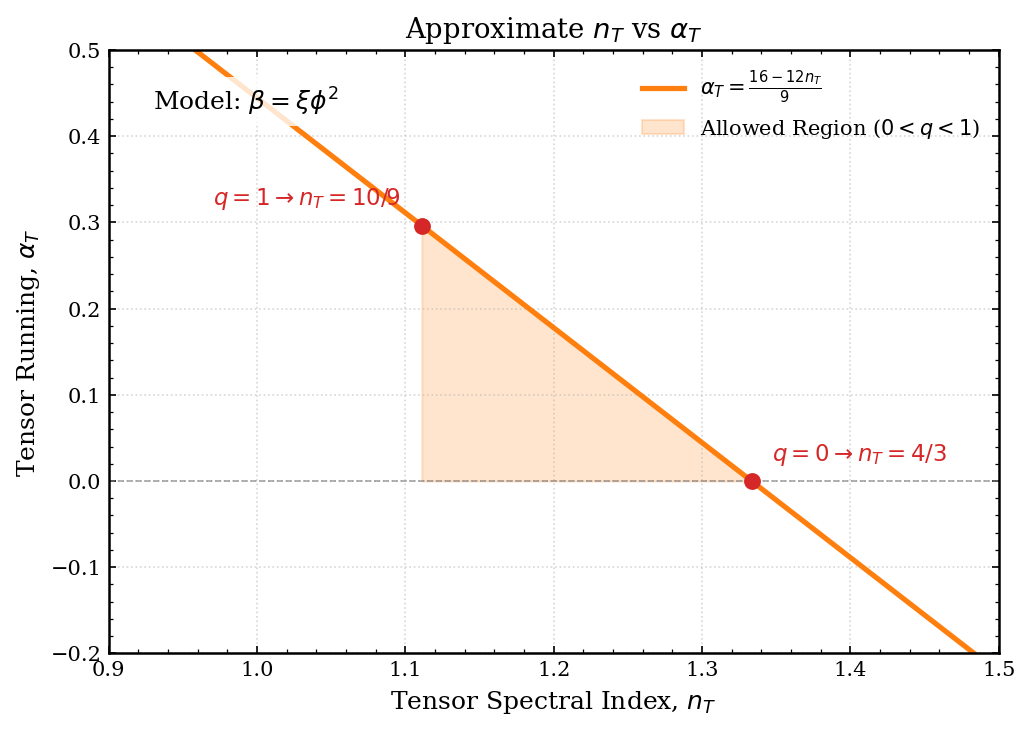}\\
        (a)
        \label{fig:plot2}
    \end{minipage}
    \begin{minipage}{0.48\textwidth}
        \centering
        \includegraphics[width=\textwidth]{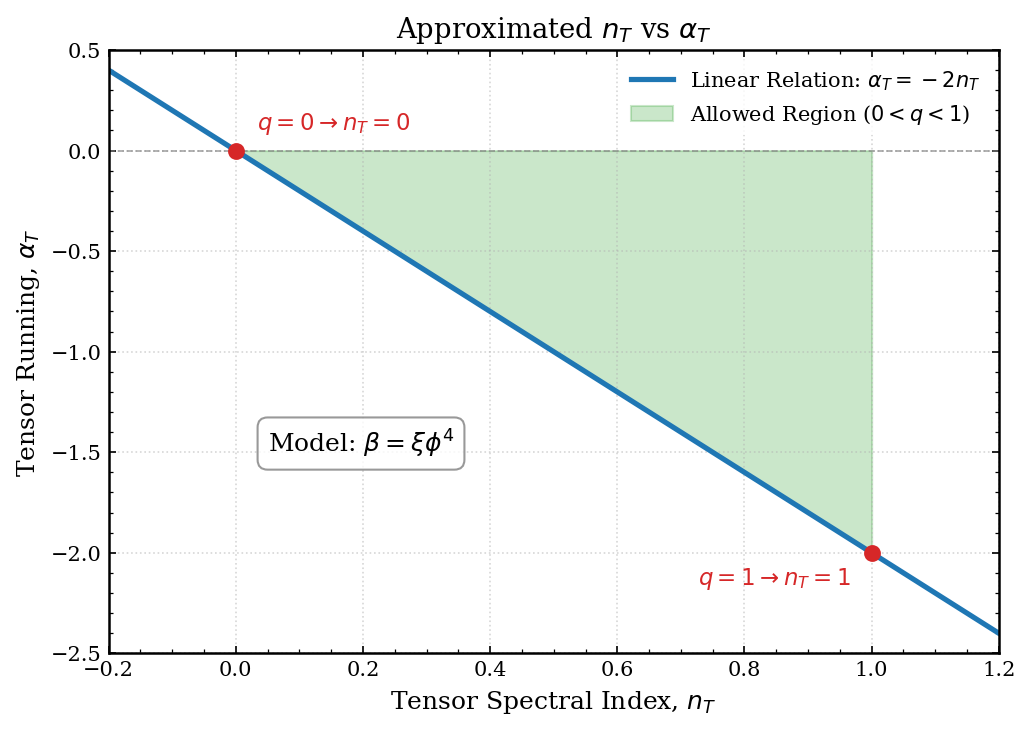}\\
        (b)
        \label{fig:plot4}
    \end{minipage}
    \caption{Small $q$ approximation to the $n_T-\alpha_T$ consistency relations for the (a) quadratic coupling $\beta=\xi \phi^2$ (b) quartic coupling $\beta=\xi \phi^4$. Approximated consistency relations for $n_T$ and $\alpha_T$ for the quadratic $\beta=\xi\phi^2$ and quartic coupling model $\beta=\xi\phi^4$. The shaded region corresponds to the physically allowed range $0<q<1$.}
    \label{fig:approximated_consistency_phi2_phi4}
\end{figure*}

Fig. \ref{fig:approximated_consistency_phi2_phi4} shows the corresponding consistency relation in the small-coupling limit. The quadratic model reduces to the approximate relation $12n_T=16-9\alpha_T$, while the quartic model reduces to $\alpha_T=-2n_T$. The shaded regions correspond to the physically allowed range $0<q<1$. Although these approximate relations are simpler than the exact expressions, they retain the essential distinction between the two models and provide a transparent representation of their observational signatures. They explicitly demonstrate that the sign of $\alpha_T$ remains model-dependent.

A future simultaneous measurement of both $n_T$ and $\alpha_T$ could therefore potentially discriminate between the quadratic and quartic forms.

\textbf{B. Evolution of Lorentz-violating tensor sector:}

\begin{figure*}
    \centering
    \begin{minipage}{0.48\textwidth}
        \centering
        \includegraphics[width=\textwidth]{"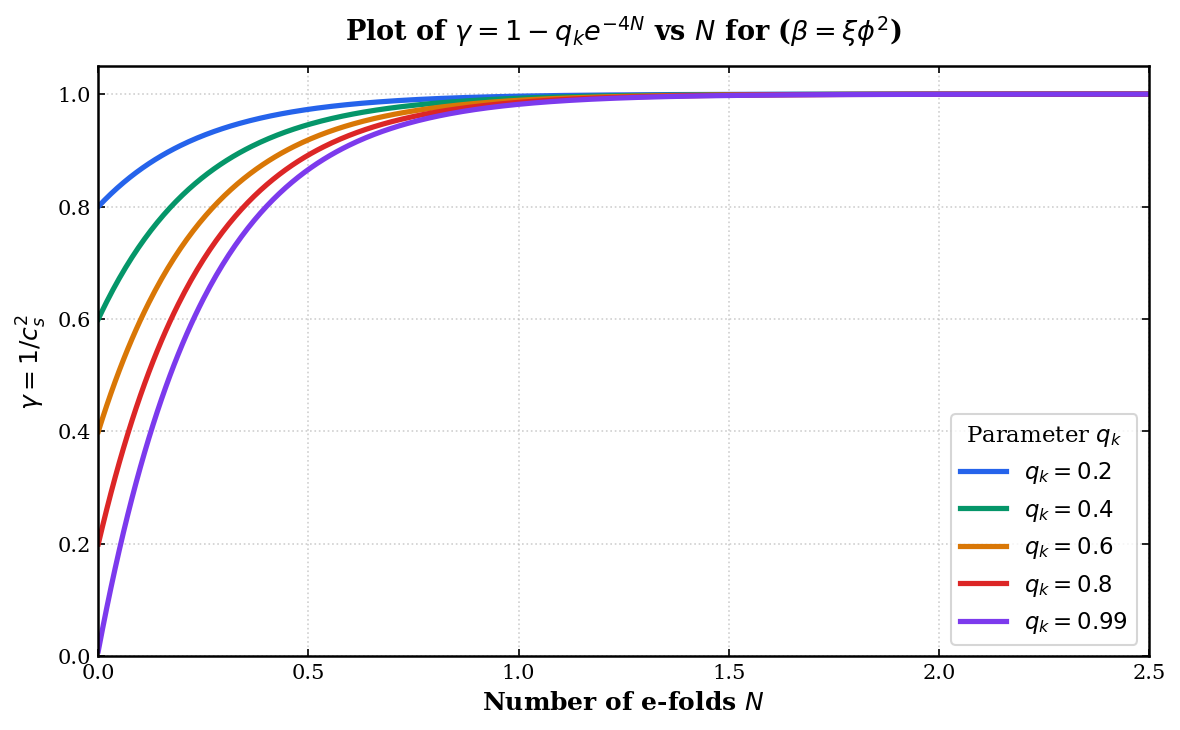"}\\
        (a) 
        \label{fig:plot6}
    \end{minipage}\hfill
    \begin{minipage}{0.48\textwidth}
        \centering
        \includegraphics[width=\textwidth]{"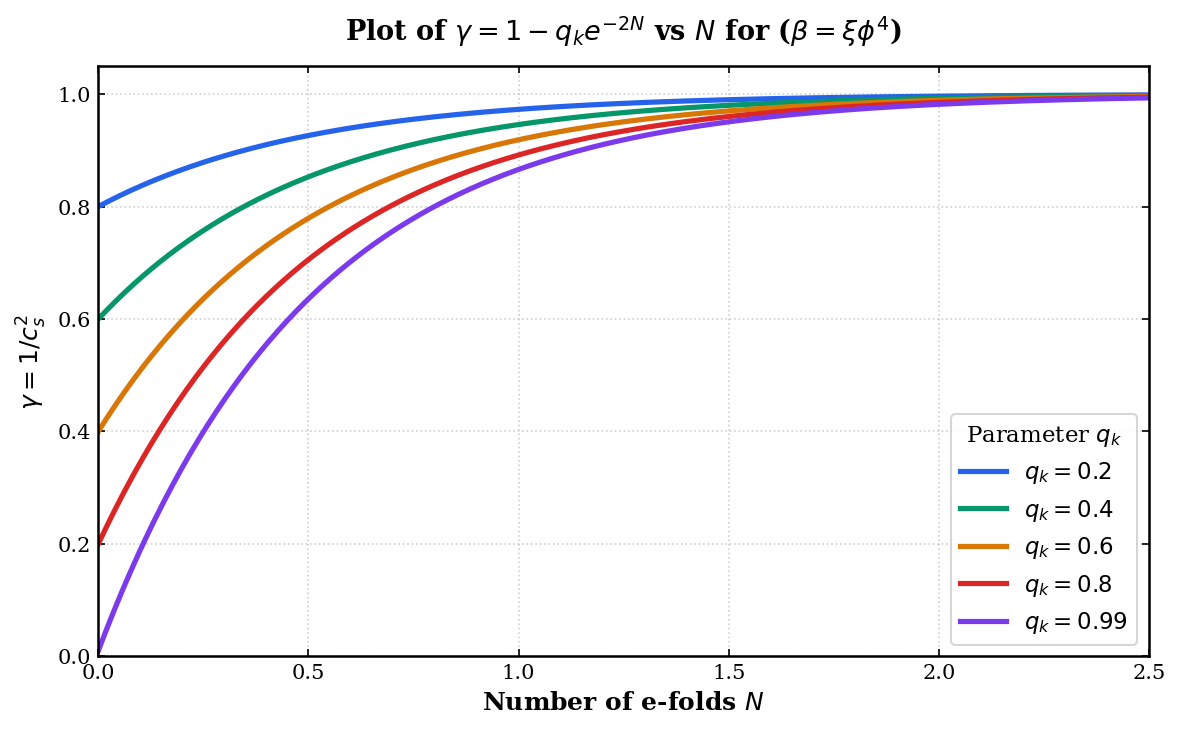"}\\
        (b) 
        \label{fig:plot7}
    \end{minipage}
    \caption{Evolution of the tensor kinetic factor $\gamma$ with the number of e-fold number $N$ for different values of the horizon crossing coupling $q_k$ for the (a) quadratic coupling $\beta=\xi \phi^2$ (b) quartic coupling $\beta=\xi \phi^4$.}
    \label{fig:gamma_vs_N}
\end{figure*}

Fig. \ref{fig:gamma_vs_N} illustrates the evolution of the tensor kinetic factor $\gamma$ as a function of the number of e-folds $N$ for different values of the horizon crossing coupling $q_k$. For smaller $q_k$, $\gamma$ remains close to unity, corresponding to a tensor sector close to the general relativistic limit. Increasing $q_k$ produces a larger initial deviation from unity and hence a stronger Lorentz-violating contribution imprint. As inflation proceeds, $\gamma$ evolves towards unity in both models, showing that the Lorentz-violating contribution becomes progressively suppressed. The initial magnitude of the Lorentz-violating coupling directly controls the strength of its imprint on tensor propagation.

An important difference between the two models is the rate at which this imprint disappears. Since $q\propto a^{-4}$ for the quadratic coupling, the Lorentz-violating contribution decreases rapidly with the expansion. In contrast, $q\propto a^{-2}$ for quartic coupling, so the Lorentz-violating contribution evolves more slowly. Consequently, the quartic model retains a measurable theoretical imprint in $\gamma$ for a large number of e-folds, whereas the quadratic model approaches the standard behaviour more rapidly. This slow convergence in the quartic case is visible in Fig. \ref{fig:gamma_vs_N} and provides a useful dynamical distinction between the two coupling functions.  

\textbf{C. Combined Early and Late-time parameter constraints:}

\begin{figure*}
    \centering
    \begin{minipage}{0.48\textwidth}
        \centering
        \includegraphics[width=\textwidth]{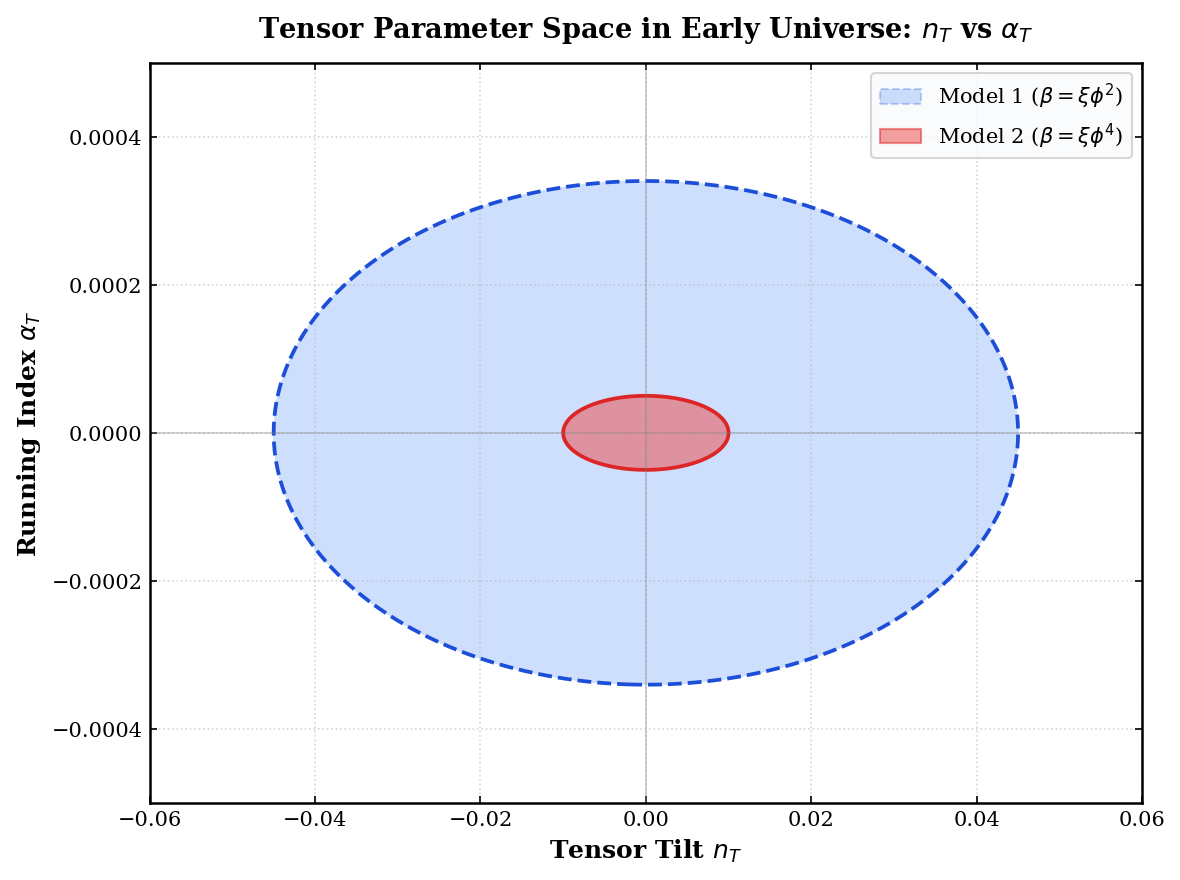}\\
        \label{fig:plot10}
    \end{minipage}\hfill
    \begin{minipage}{0.48\textwidth}
        \centering
        \includegraphics[width=\textwidth]{"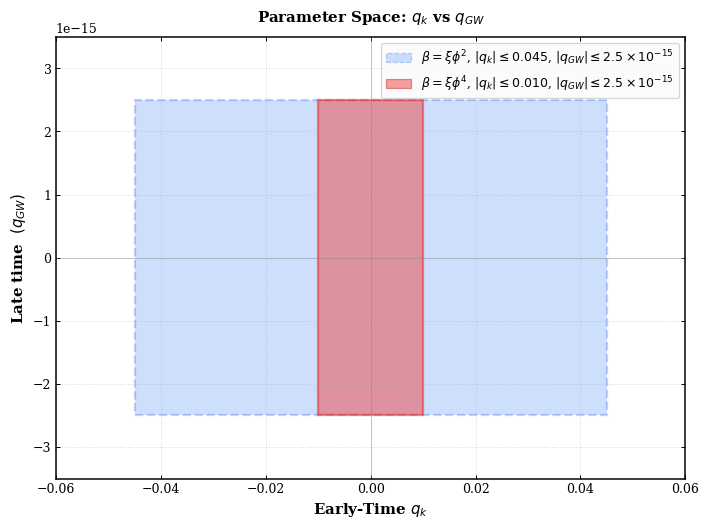"}\\
        (b) Late time
        \label{fig:plot12}
    \end{minipage}
    \caption{Combined early and late-time constraints on the tensor-sector coupling for the quadratic and quartic models. (a) Primordial parameter space in the $(n_T-\alpha_T)$ plane obtained from the allowed range of $q_k$ at horizon crossing. (b) Comparison of the horizon-crossing coupling $q_k$ with the late-time coupling $q_{GW}$ constrained by the GW propagation speed. The shaded regions indicate corresponding allowed parameter ranges.}
    \label{fig:parameter_space_summary}
\end{figure*}

Fig. \ref{fig:parameter_space_summary} summarizes the parameter-space constraints from primordial and late-time tensor observables. The early-time panel displays the allowed region in the $(n_T,\alpha_T)$ plane obtained from the constraints on $q_k$ at the horizon crossing. The quadratic and quartic models occupy distinct regions, reflecting their different consistency relations. The quadratic model allows a somewhat larger primordial range $|q_k|\lesssim 0.045$, whereas the quartic model is restricted to approximately $|q_k|\lesssim 0.010$. Thus, primordial tensor observables impose stronger restrictions on the quartic coupling model.  The late-time panel compares the primordial coupling $q_k$ with the late-time coupling $q_{GW}$ constrained by the gravitational-wave propagation speed at the epoch of GW170817. The stringent bound on $q_{GW}$ is many orders of magnitude stronger (extremely close to zero) than the primordial bounds for both models. However, $q_k$ and $q_{GW}$ refer to different cosmological epochs and should not be identified automatically. 

Taken together, Figs. \ref{fig:nT_vs_q}-\ref{fig:parameter_space_summary} demonstrate that the two coupling choices are not merely alternative parametrizations of the same tensor phenomenology. Their different field dependence produces different background evolution, different decay rates of the effective Lorentz-violating contribution, distinct $n_T-\alpha_T$ consistency relations, and different parameter-space constraints. The tensor tilt and its running therefore provide complementary primordial probes, while gravitational wave propagation speed supplies a powerful late-time constraint.

\onecolumngrid
\vspace{1em}

\FloatBarrier

\twocolumngrid

\section{Conclusion}
In this work, we investigated the tensor-sector signatures of Lorentz-violating inflation with non-canonical scalar kinetics in Einstein-\AE{}ther gravity. We considered a scalar sector containing a higher-order kinetic correction $K_3X^2$ and two representative field-dependent \AE{}ther couplings, $\beta\sim\phi^2$ and $\beta\sim \phi^4$. We first examined the modified inflationary background and showed that the Lorentz-violating and standard slow-roll regimes lead to qualitatively different background dynamics for the two coupling forms. We further studied the zero potential case and obtained exact background solutions. The quadratic coupling leads to super inflationary evolution, whereas the quartic coupling admits an exact de Sitter expansion with non-trivial scalar field dynamics. We find that the Lorentz-violating phase exists for the two coupling models:$\beta\sim\phi^2$ and $\beta\sim \phi^4$ within the parameter ranges $1/226<\xi<1/16$ and $1/648\pi <\xi< 1/324$, respectively. Thus, the analysis demonstrates that accelerated expansion can arise from the interplay between Lorentz-violating coupling and non-canonical kinetic dynamics even in the absence of an explicit inflaton potential.

The principal focus of the work was the primordial tensor sector. We derived the tensor quadratic action and the corresponding tensor propagation speed and power spectrum. In the constant-$\gamma$ limit, the quadratic coupling produces a blue tensor spectrum, while the quartic coupling gives a scale-invariant spectrum. Allowing the Lorentz-violating parameter to evolve introduces corrections to the tensor tilt and its running. For the quadratic coupling model, the tensor observables are found to be $n_T=(4-2q)/(3-q)$ and $\alpha_T=8q(1-q)/(3-q)^3$. While for the quartic coupling model the observables found are $n_T=q$ and $\alpha_T=-2q(1-q)$.  Most importantly, eliminating the effective coupling parameter yields model-independent consistency relations between $n_T$ and $ \alpha_T$: $\alpha_T=2(3n_T-4)(n_T-1)(n_T-2)$ for $\beta\sim\phi^2$ and $\alpha_T=-2n_T(1-n_T)$ for $\beta\sim \phi^4$. These relations provide distinctive observational signatures and show that the tensor tilt and its running are not independent observables in the present framework. In both cases, the tensor propagation speed $c_T$ departs from unity and is constrained by observations of GW170817/GRB170817A. 

We then translated projected constraints on $n_T$ and $\alpha_T$ into bounds on the effective Lorentz-violating coupling at the horizon crossing. Future measurements of primordial tensor observables could constrain the coupling parameter at horizon crossing to lie within $|q_k|\lesssim4.5\times 10^{-2}$, thereby placing the coupling within a well-defined range during inflation. Future experiments such as LiteBIRD, CMB-S4, DECIGO and BBO are expected to test much weaker constraints by measuring the tensor spectral index $n_T$ and its running $\alpha_T$ \cite{LiteBIRD:2023ptep,LiteBIRD:2024jcap, CMBS4:2016book,CMBS4:2017tech,DECIGO:2011cqg,DECIGO:2021ptep, BBO:2005prd,BBO:2006cqg,BBO:2006prd}. The resulting bounds are substantially weaker than the stringent late-time constraint obtained from the propagation speed of gravitational waves, which constrains $|q_{GW}|\lesssim 2.5\times 10^{-15}$. However, the latter constraints the coupling $q_{GW}$ at the epoch of GW170817, whereas the primordial observables constrain $q_k$ at the horizon crossing. Therefore, identifying the two parameters requires the additional assumption that the effective tensor-sector coupling remains approximately unchanged between inflation and the late universe.

This distinction is important for interpreting the GW170817 bound. Rather than assuming an automatic equality $q_{GW}=q_k$, the late-time observation should be regarded as a direct constraint on the present-day tensor-sector coupling. A more general treatment can be formulated through an evolution factor relating the two epochs, which would allow the primordial and late-time constraints to be connected without assuming a constant Lorentz-violating coupling.

Overall, our results demonstrate that the combination of non-canonical scalar dynamics and field-dependent Lorentz violation can leave potentially observable and model-dependent signatures in the primordial tensor sector. The $(n_T-\alpha_T)$ consistency relations provide a particularly clean way of distinguishing the quadratic and quartic coupling scenarios. Future measurements of the tensor spectral index, its running and primordial gravitational wave backgrounds could therefore provide complementary probes of Lorentz-violating inflation. A comparative analysis of the tensor sector observables and parameter space constraints reveals that quadratic and quartic coupling models yield distinct phenomenological signatures. It can be seen from Fig.\ref{fig:gamma_vs_N} that $\gamma$ approaches unity at the beginning of inflation $(N\sim0)$ for smaller values of the coupling parameter $q_k$, corresponding to the general relativistic limit, while larger values of $q_k$ lead to greater deviation of $\gamma$ from unity, hence a stronger Lorentz-violation effect for both coupling models. As the number of e-folds increases, $\gamma$ gradually converges towards unity, with convergence occurring more rapidly in the quadratic model than in the quartic coupling model. This indicates that the Lorentz-violation effects decrease as inflation proceeds.

Although the present analysis has focused on the background evolution and tensor perturbations, several important directions remain to be explored. First, it would be valuable to study the scalar sector and develop a complete phenomenological framework for Lorentz-violating inflation with non-canonical kinetics. In particular, an analysis of scalar perturbations, the scalar sound speed and the associated non-Gaussian signatures would provide additional observable tests of the theory. It will also be important to investigate the full stability conditions and to investigate the reheating phase, thereby connecting the inflationary epoch to the subsequent evolution of the universe. A combined analysis of scalar and tensor observables would provide a more comprehensive assessment of the observational viability of the model. Finally, detailed forecasts for future primordial gravitational wave detectors will be useful in probing the tensor signatures and the consistency relations identified in this work.

\FloatBarrier 
\bibliographystyle{unsrt}
\bibliography{references}
\end{document}